\documentclass[twocolumn,showpacs,amsmath,amssymb,prb,reprint,superscriptaddress,floatfix]{revtex4-1}
\usepackage{bbm,graphicx,color,comment,txfonts,dsfont}
\usepackage[bookmarks=true,colorlinks,citecolor=blue,urlcolor=blue]{hyperref}
\usepackage{dsfont}
\usepackage{feynmf}
\usepackage{dcolumn}

\begin{document}
\title{Nodal points of Weyl semimetals survive the presence of moderate disorder}
\author{Michael Buchhold}
\affiliation{Department of Physics and Institute for Quantum Information and Matter, California Institute of Technology, Pasadena, CA 91125, USA}
\author{Sebastian Diehl}
\affiliation{Institute for Theoretical Physics, Universit\"at zu K\"oln,  D-509237 K\"oln, Germany}
\author{Alexander Altland}
\affiliation{Institute for Theoretical Physics, Universit\"at zu K\"oln,  D-509237 K\"oln, Germany}

\begin{abstract}
In this work we address the physics of individual three dimensional Weyl nodes subject to a moderate concentration of disorder. Previous analysis indicates the presence of a quantum phase transition below which  disorder becomes irrelevant and the integrity of sharp nodal points of vanishing spectral density is preserved in this system. This statement appears to be at variance with the inevitable presence of  statistically rare fluctuations which cannot be considered as weak and must have strong influence on the system's spectrum, no matter how small the average concentration. We here reconcile the two pictures by demonstrating that rare fluctuation potentials in the Weyl system generate a peculiar type of resonances which carry spectral density in any neighborhood of zero energy, but never at zero. In this way, the vanishing of the DoS for weak disorder survives the inclusion of rare events. We demonstrate this feature by considering three different models of disorder, each emphasizing specific aspects of the problem: a simplistic box potential model, a model with Gaussian distributed disorder, and one with a finite number of $s$-wave scatterers. Our analysis also explains why the protection of the nodal DoS may be difficult to see in simulations of finite size lattices. 
\end{abstract}
\date{\today}
\maketitle

\section{Introduction} 
Weyl materials is the overarching terminology for three dimensional quantum matter containing an even number of separate linearly dispersive  Dirac cones  in the Brillouin zone. The presence of these nodes manifests itself in a variety of effects, including the formation of surface Fermi arc states \cite{Fermiarc2}, chiral magnetotransport \cite{Yan2017}, and nonlinear optical response \cite{NLOR1,NLOR2}. These and various other unconventional phenomena have attracted a lot of recent experimental~\cite{Xu613, Inoue2016, Weng,Belo2016,   Xu2015z,Xu2016z} and theoretical~\cite{VishRev,Surface1, Surface2, SyzranovRev,
Fermiarc1, ChiralAnThe1} attention.

Individual Weyl cones are protected by topology. While their nodal centers can be moved  in the Brillouin zone, the definite chirality of individual nodes prevents the opening of a spectral gap. Topology does, however, not protect the Weyl cones from becoming 'soft', i.e. a smoothening of the linear spectrum and the replacement of the singular nodes by a continuum of states with finite spectral density. Especially in the vicinity of the band touching points, the spectrum  responds sensitively towards perturbations due to, e.g., random impurities or defects and this poses the question if, or under what circumstances, the spectral integrity of the nodes  is preserved away from the limit of a pristine Weyl Hamiltonian. 

This question motivated a large number of studies addressing the physics of Weyl materials subject to static, potential disorder, which lead to an ongoing and partly controversial debate. Two fundamentally different pictures have been drawn, each supported by different analytical and numerical theories: perturbative renormalization group theory\cite{Fradkin86b,
Fradkin86a, Sbierski2017,GurarieWeyl,Goswami2011,Roy2014,Roy2016, Syzranov2016b,GurarieWeyl2, GurarieWeyl3} in $d=2+\epsilon$ dimensions, self-consistent Born approaches\cite{Fradkin86b, Fradkin86a} and a variational nonlinear sigma
model approach\cite{Altland2015, Altland2016} predict the semimetallic phase to be robust against weak disorder. Within these approaches, weak disorder is irrelevant in a renormalization group sense, and only for concentrations above a critical threshold, $K_c$, the system will enter a metallic phase with globally non-vanishing density of states. These phases are separated from each other by a quantum critical point, where the average zero energy density of states (DoS) $\langle \nu(0)\rangle$ --- zero in one phase, finite in the other --- serves as an order parameter.  The observation of a quantum critical point has been confirmed numerically \cite{Sbierski2015,Sbierski2014,Shapourian,Soumya,Kobayashi2014,Shang2016,Slager2017,Roy2018} and the corresponding scaling exponents have been determined. 

On the other hand, it has been argued that the vanishing density is at variance with the presence of  bound states generated by rare disorder configurations~\cite{Nandkishore:2014aa}. A variational approach for the DoS indeed showed that there exist 'optimal' disorder configurations, which are able to bind quantum states at zero energy. It is natural to expect that such states generate finite spectral density at zero energy. In the same ballpark of theoretical approaches there is work\cite{Balatsky2014} on models with few strong isolated impurities, which likewise has demonstrated the existence of bound states. This picture is supported by  numerical studies of lattices with finite disorder correlation length~\cite{Pixley2016a,Pixley2016b,Wilson2016,Pixley2015c,Ostrovsky2017}.  

The two different families of approaches outlined above are at variance in that a finite zero energy DoS, albeit not categorically ruling out a phase transition \cite{GurarieWeyl2, GurarieInst}
, would not sit comfortably with quantum criticality at a finite disorder concentration. 
In this work we reconcile the two pictures and demonstrate that the undeniable presence of rare events at any disorder concentration does not compromise the integrity of the nodes~\cite{BuchholdLetter}. We will demonstrate that fluctuations may generate spectral weight everywhere, except at zero energy. Specifically, in cases where the centers of impurity resonances approach zero energy, their width narrows, and a `screening' effect keeps the nodal DoS  pinned to zero. 
We will demonstrate this behavior for three different models, each emphasizing different aspects of the problem. The first is just a spherically symmetric potential well as a cartoon version of an isolated potential fluctuation. This model is oversimplifying  but affords a rigorous analytic solution demonstrating key features of the general situation. The second model describes the disorder via a Gaussian distributed potential. Focusing on isolated rare fluctuations of the latter, we apply instanton calculus to demonstrate that the vanishing of the zero energy DoS did not rely on artificial aspects of the first model. The third model describes a dilute system of strong impurities, and in this way includes the effect of impurity correlations into the analysis.

The notorious stability of the nodes is not backed by any mechanism involving symmetry, or even topology. Rather, what makes the problem distinct from rare fluctuations of a Schr\"odinger operator is that the eigenstates of a Weyl Hamiltonian never show exponential behavior. This implies an extensive level of hybridization between the region of a strong potential fluctuation with the outside. At zero energy the absence of states in the unperturbed problem prohibits the existence of zero energy states, including in the presence of isolated strong potential modulations. In the rest of the paper, we will demonstrate this behavior explicitly for our three different setups. In section Sec.~\ref{SingleImp} we consider the prototypical setup of a spherically symmetric box potentials, for which exact expressions for the DoS and its distribution can be obtained.  In Sec.~\ref{SUSY}, we discuss the instanton approach to rare event formation in Gaussian distributed disorder potentials, and in Sec.~\ref{MultImp} a $T$-matrix  approach is applied to obtain the DoS for a system with multiple impurities and band curvature. We conclude in section Sec.~\ref{sec:Conclusion}.

\section{Phase shift and DoS in a single, spherical box potential}\label{SingleImp}

The spectrum of a single node in a Weyl semimetal subject to a potential $V_x$ is obtained by diagonalization of the Hamiltonian
\begin{align}\label{EqHam1}
\hat H = -i v_0\sigma_i\partial_i+V_x,
\end{align}
where an isotropic nodal velocity $\nu_0$ is assumed, and $\sigma_i$ are the Pauli matrices. {\color{black} (For a brief discussion of more general sources of disorder, see section~\ref{sec:Conclusion}.)} For notational simplicity, we will set $v_0=1$ unless stated otherwise. In this section, we consider the case of a spherically symmetric box potential $V(\vec{r})=\lambda\ \theta(r_0-|\vec{r}|)$, of radius $r_0$ and depth, $\lambda>0$ (due to the symmetry of the spectrum around zero, it suffices to consider positive depths.) While this can be no more than a cartoon of a realistic potential fluctuation, textbook methods can be applied to a solution which exhibits many of the general characteristics of the problem. 

As discussed in more detail below, a feature distinguishing the Hamiltonian~\eqref{EqHam1} from the corresponding three-dimensional Schr\"odinger Hamiltonian is that a box potential does not bind states, except for at special configurations $\lambda r_0=n\pi$ with $n\in \mathbb{N}$. At these 'magical' values, the system supports a single bound state at zero energy. At all other values, the presence of the potential leads to resonances at finite energy, i.e. states with a wave function that is extended over the whole system (and not bound) but resemble bound states at short distances. Such resonances modify the density of states, 
\begin{align}\label{EqHam63}
\nu(\omega)=\nu_0(\omega)+\delta\nu(\omega),
\end{align}
where $\nu_0(\omega)$ is the DoS of the clean Weyl system and $\delta\nu(\omega)$ 
the modification due to the box potential. Absent bound states, the 
 Friedel sum rule\footnote{Actually, the application of the Friedel sum rule to the present problem is not totally innocent. The traditional derivation\cite{Friedel1952} is based on an argument which assumes the entire system to be put into a large box of finite extension. However, topology implies that a single Weyl node cannot be put in a box. And any box setup containing an even number of nodes would come with inter-node scattering at the boundaries which we absolutely want to avoid. Fortunately, there exists an alternative derivation of the sum rule\cite{LangerSum} which does not require finite confinement volumes, nor other requirements at odds with the topology of the Weyl node.} 
requires
\begin{align}\label{EqHam64}
\int_\omega \delta\nu(\omega)=0 
\end{align} 
This states that the DoS accumulated by resonances at some energy $\omega_0$ is exactly pulled away from other states in the spectrum, and no net DoS is generated by the potential. The key result of this section will be that Eq.~\eqref{EqHam64} is always fulfilled in such a way that 
\begin{align}\label{EqHam65}
\delta\nu(\omega=0)=0.
\end{align}
At the magical values, the zero energy density of states, $\delta\nu(\omega)$ \emph{is} modified by the $\delta$-function representing the bound state. However, this spectral weight is `screened' by an equally singular counterweight in infinitesimal proximity of zero. As a consequence, any finite region containing zero does not harbor spectral weight, including at the magical configurations.  

A different way of formulating the result is to consider the parameters $\lambda,r_0$ statistically distributed according to a distribution $P(\lambda,r_0)$, $\lambda, r_0\in \mathbb{R}^+$. The magical values $\lambda r_0=n\pi$ define a discrete subset of zero measure. Performing an average over all possible impurity configurations $\langle\nu(\omega)\rangle=\int_{\lambda,r_0}\nu(\omega)$, the average DoS at zero energy thus effectively remains zero, both in a realization specific and in a statistical sense.

\subsection{Eigenstates and phase shift in a box potential}

The eigenfunctions of the Weyl problem subject to a spherical box potential have been determined in several works\cite{Greiner, Friedel,Nandkishore:2014aa}. In this section we briefly review their derivation and their essential properties. These results will play a role as building blocks in the later parts of the analysis. 

Due to the spherical symmetry of the problem, the eigenfunctions $\psi_{\omega,\kappa,m_j}({\bf r})=\langle {\bf
r}|\omega,\kappa,m_j\rangle$ of energy $\omega$ can be organized in multiplets labeled by the total half-integer angular momentum
$j=\kappa-\frac{1}{2}$, $\kappa\in \Bbb{N}$, and angular momentum orientation $m_j$.
In spherical coordinates, the box Hamiltonian then reads as
\begin{align}
\label{EqHam66}
H=\frac{i}{r}\vec{\sigma}\cdot\vec{r}\left(\partial_r-\frac{\vec{\sigma}\cdot\vec{L}}{r}\right)+\lambda\theta(r_0-r).    
\end{align}
We represent the eigenfunctions as 
\begin{align}
\label{EqHam67}    
\psi_{\omega,\kappa,m_j}(r,\theta,\phi)&=f_{\omega,\kappa,-}(r)\chi_{j,-,m_j}(\theta,\phi)+i f_{\omega,\kappa,+}(r)\chi_{j,+,m_j}(\theta,\phi),\ \ \ \ \ 
\end{align}
where the angular factors split the total angular momentum, $j$,  into a spin component $s=\pm\frac{1}{2}$ and an orbital momentum $l=j\pm\frac{1}{2}=\kappa,\kappa-1$. These functions are defined by the relations
\begin{align}
\label{EqHam68}    
 \vec{J}^2\chi_{j,\pm,m_j}&=j(j+1)\chi_{j,\pm,m_j}, \ \ J_z\chi_{j,\pm,m_j}=m_j\chi_{j,\pm,m_j}, \\
 \vec{L}^2\chi_{j,\pm,m_j}&=(j\pm\tfrac{1}{2})(j\pm\tfrac{1}{2}+1)\chi_{j,\pm,m_j},\label{EqHam69}\\
\tfrac{\vec{\sigma}\cdot\vec{r}}{r}\chi_{j,\pm,m_j}&=-\chi_{j,\mp,m_j}.\label{EqHam70}
\end{align}
Due to Eqs.~\eqref{EqHam68}-\eqref{EqHam69} they are pairwise orthogonal and we choose them to be normalized with respect to the spherical integration
\begin{align}
\label{EqHam71}    
\int  \chi^*_{j,\alpha,m_j}\chi_{j',\beta,m_j'} \sin\theta d\phi d\theta=\delta_{j,j'}\delta_{\alpha,\beta}\delta_{m_j,m_j'}.
\end{align}
Combining Eqs.~\eqref{EqHam67}-\eqref{EqHam71}, we obtain the eigenequations for the radial wave functions
\begin{align}
\label{EqHam72}    
\left(\partial_r-\frac{\kappa-1}{r}\right)f_{\omega,\kappa,-}&=\left(\lambda\theta(r_0-r)-\omega\right)f_{\omega,\kappa,+},\\
\left(\partial_r+\frac{\kappa+1}{r}\right)f_{\omega,\kappa,+}&=-\left(\lambda\theta(r_0-r)-\omega\right)f_{\omega,\kappa,-}.\label{EqHam73}
\end{align}
They can be solved independently for $r<r_0$ and $r>r_0$, yielding 
\begin{align}
\label{EqHam74}   
f_{\omega,\kappa,+}= \left\{\begin{array}{ll}\frac{A^<_{\omega,\kappa} J_{\kappa+1/2}((\omega-\lambda)r)+B^<_{\omega,\kappa}Y_{\kappa+1/2}((\omega-\lambda)r)}{\sqrt{(\omega-\lambda)r}}& \text{ for } r<r_0\\ \frac{A^>_{\omega,\kappa} J_{\kappa+1/2}(\omega r)+B^>_{\omega,\kappa} Y_{\kappa+1/2}(\omega r)}{\sqrt{\omega r}}& \text{ for } r>r_0\end{array}\right. 
 \end{align}
in terms of the Bessel functions of the first and second kind $J, Y$. Continuity at $r=0$ requires $B^<=0$ and we can set $A^<=1$. The second radial function $f_{\omega,\kappa,-}$ can be obtained by combining \eqref{EqHam73} and \eqref{EqHam74} and continuity of both functions at $r=r_0$ determines the coefficients $A^>, B^>$ as
\begin{align}
\label{EqHam75}    
A^>_{\omega,\kappa}&=\tfrac{ Y_{\kappa -\frac{1}{2}}(r_0\omega) J_{\kappa +\frac{1}{2}}(r_0 (\omega-\lambda))-Y_{\kappa +\frac{1}{2}}(r_0\omega) J_{\kappa -\frac{1}{2}}(r_0 (\omega-\lambda))}{\sqrt{\frac{\omega-\lambda}{\omega}} \left(J_{\kappa +\frac{1}{2}}(\omega r_0) Y_{\kappa -\frac{1}{2}}(r_0\omega)-J_{\kappa -\frac{1}{2}}(r_0\omega) Y_{\kappa +\frac{1}{2}}(r_0\omega)\right)},\\
B^>_{\omega,\kappa}&=\tfrac{ J_{\kappa +\frac{1}{2}}(r_0\omega) J_{\kappa -\frac{1}{2}}(r_0 (\omega-\lambda))-J_{\kappa -\frac{1}{2}}(r_0\omega) J_{\kappa +\frac{1}{2}}(r_0 (\omega-\lambda))}{\sqrt{\frac{\omega-\lambda}{\omega}} \left(J_{\kappa +\frac{1}{2}}(\omega r_0) Y_{\kappa -\frac{1}{2}}(r_0\omega)-J_{\kappa -\frac{1}{2}}(r_0\omega) Y_{\kappa +\frac{1}{2}}(r_0\omega)\right)}.\label{EqHam76}
\end{align}
This choice of $A^>, B^>$ leads to a continuous behavior of $f_{\omega,\kappa,\pm}$ as a function of $\omega$ and reproduces correctly the limiting cases $\omega\rightarrow0,\lambda$. 

In the  limit $r\rightarrow \infty$ the scattering wave functions approach the solutions of the clean Weyl Hamiltonian $(\lambda=0)$, modified by a scattering phase shift $\delta_\kappa(\omega)$, i.e.
\begin{align}
\label{EqHam77}
f_{\omega,\kappa,-}&\overset{r\rightarrow\infty}{\rightarrow}\frac{1}{r}\sin(\omega r-\frac{\kappa\pi}{2}+\delta_k(\omega)).  
\end{align}
Using the asymptotic behavior of the Bessel functions for large arguments, one can write this phase shift as
\begin{align}
\label{EqHam78} 
\tan\delta_\kappa(\omega)=\tfrac{J_{\kappa-\frac{1}{2}}(r_0\omega)J_{\kappa+\frac{1}{2}}(r_0(\omega-\lambda))-J_{\kappa-\frac{1}{2}}(r_0(\omega-\lambda))J_{\kappa+\frac{1}{2}}(r_0\omega)}{J_{\kappa-\frac{1}{2}}(r_0(\omega-\lambda))Y_{\kappa+\frac{1}{2}}(r_0\omega)-J_{\kappa+\frac{1}{2}}(r_0(\omega-\lambda))Y_{\kappa-\frac{1}{2}}(r_0\omega)}.   
\end{align}
 
Equations \eqref{EqHam75}-\eqref{EqHam78} describe the scattering wave functions of the Hamiltonian~\eqref{EqHam66}, i.e. its extended, non-normalizable eigenfunctions. Away from the special point $\omega=0$, no localized eigenstates (bound states)  exist and all eigenstates are extended\cite{Pieper1969, Greiner}. At $\omega=0$, however, Eqs.~\eqref{EqHam72} and \eqref{EqHam73} decouple and one may find localized (and square-integrable) solutions of the form (for $r>r_0$)
\begin{align}
\label{EqHam79ex}
f_{0,\kappa,-}=0, \ \ f_{0,\kappa,+}=r^{-\kappa-1}.    
\end{align}
Continuity of the solutions with definite $\kappa$ at $r=r_0$ requires
\begin{align}
\label{EqHam80ex}    
&J_{\kappa-\frac{1}{2}}(-\lambda r_0)\overset{!}{=}0,&\\
&\text{ e.g. for }\left\{ \begin{array}{cc}\kappa=1: &\sin(r_0\lambda)=0\\
\kappa=2:&\tan(r_0\lambda)=\lambda\end{array}\right.&\nonumber .
\end{align}
For $\kappa=1$, bound states exist at the aforementioned 'magical' values
$r_0\lambda=n\pi$, $n\in \mathbb{N}$. For $\kappa>1$ bound states have to fulfill a
transcendental equation; in contrast to $\kappa=1$ these consistency equations are not solved by equidistantly spaced
values of the parameter $r_0\lambda$. Furthermore, the minimal value
$(r_0\lambda)_{\text{min}}$ for which bound states occur increases approximately
linear in $\kappa$, requiring stronger bound state potentials for larger angular
momentum states. At the magical values, the scattering phase performs a discontinuous
jump of $\Delta\delta=\pi$ when passing $\omega=0$, and in this way indicates the presence of
a bound state.

The scattering phase shift $\delta_\kappa(\omega)$ at energy $\omega$ is related to the change in the density of states $\delta\nu(\omega)=\nu(\omega)-\nu_0(\omega)$ due to the presence of the box through the Friedel sum rule
\begin{align}
\delta\nu(\omega)&=\frac{2}{\pi}\sum_{\kappa=1}^\infty\kappa\partial_{\omega}\delta_{\kappa}(\omega),\label{EqHam83a}
\end{align}
where we used the $2\kappa$-fold degeneracy of the solutions of fixed $\kappa$.
In the remainder of this section, we will show that this DoS vanishes
close to $\omega=0$ when averaged
over any continuous probability distribution for $\lambda$ and $r_0$.

\subsection{Density of states of the Weyl particle in a box potential}
We are now in the position to analyze the DoS of the Weyl particle on the basis of  Eq.~\eqref{EqHam83a}. Without loss of generality, we simplify the notation by setting $r_0=1$  and discuss contributions $\delta\nu_\kappa(\omega)\equiv\frac{2\kappa}{\pi}\partial_\omega\delta_{\kappa}(\omega)$ from different angular momentum states $\kappa$ separately. The configurations stabilizing a bound state of given $\kappa$ are denoted by $\lambda_{\kappa,c}$, e.g. $\lambda_{1,c}=n\pi$.

\begin{figure}
	\includegraphics[width=\linewidth]{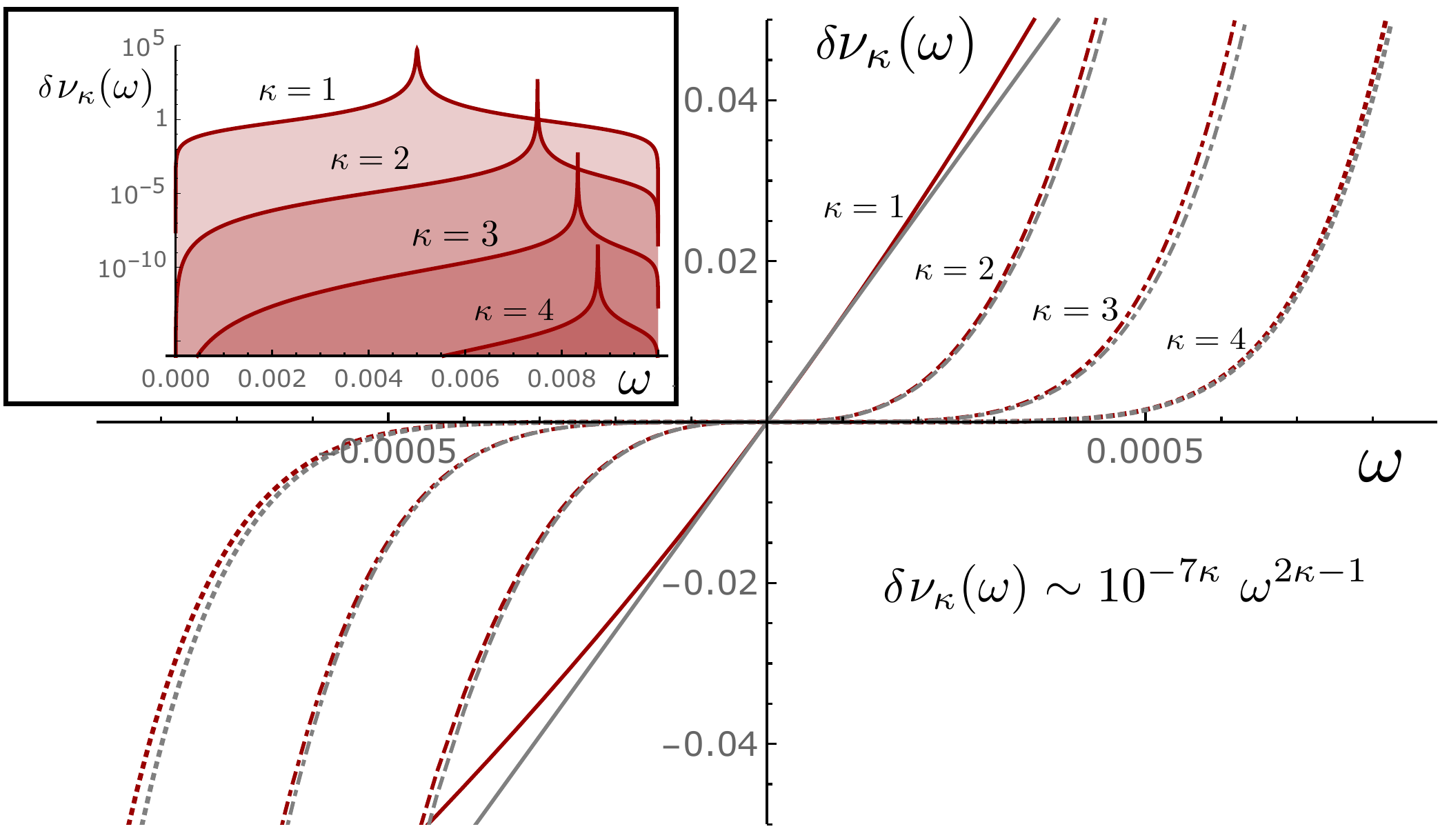}
	\caption{Density of states, $\delta\nu_{\kappa}(\omega)$ for $\kappa=(1,2,3,4)$ and for a deviation from the lowest bound state potential configuration $\Delta_\kappa=0.01$. The central figure shows the numerical values of the DoS (red lines) in the regime $-10^{-3}<\omega<10^{-3}$, where it is well approximated by a $\kappa$-dependent monomial $\delta\nu_\kappa(\omega)\sim \omega^{2\kappa-1}$ (grey lines). The numerical evaluation shows that for increasing $\kappa$, the correction to the bare DoS is suppressed by a factor $10^{-7(\kappa-1)}$. The DoS in the central figure is rescaled accordingly. The inset demonstrates the accumulation of spectral density in the regime $0<\omega<\Delta_\kappa$ on a logarithmic scale, again for $\Delta_\kappa=0.01$. Besides the evolution of the peak towards larger values of $\omega$ as a function of $\kappa$, it reveals the strong suppression of $\delta\nu_{\kappa}(\omega)\sim 10^{-7(\kappa-1)}$ for increasing $\kappa$.}
	\label{fig1}
\end{figure}
Small deviations from a magical potential configuration $\Delta_\kappa\equiv\lambda-\lambda_\kappa\neq0$ lead to resonances in the DoS $\delta\nu_\kappa(\omega)$ at frequencies $\omega\sim \Delta_\kappa$. This resonant behavior manifests itself in a strong increase of the DoS, $\delta\nu_\kappa(\omega\sim\Delta_\kappa)>0$. At the same time, in the absence of bound states the Friedel sum rule~\eqref{EqHam64} requires a vanishing integral of the impurity DoS. Due to the independence of different angular momentum sectors, this sum rule must in fact hold for individual $\kappa$,  
\begin{align}
\label{EqHam86}
\int_\omega \delta\nu_\kappa(\omega)\overset{!}{=}0.   
\end{align}
The analysis of $\delta\nu_\kappa(\omega)$ at $|\omega|\ll |\Delta_{\kappa}$ shows that  the redistribution of spectral weight happens in such a way that $\delta\nu_\kappa(0)=0$ for all $\Delta_{\kappa}\neq0$. In fact, a careful (but lengthy) expansion of the scattering phase shift $\delta_{\kappa}(\omega)$  yields the scaling estimate
\begin{align}
\label{EqHam87}
\delta\nu_\kappa(\omega)\overset{|\omega|\ll |\Delta_\kappa|}{\sim} \text{sign}(\Delta_{\kappa})\omega^{2\kappa-1},    
\end{align}
which is very well illustrated by the numerical evaluation of the DoS in Fig.~\ref{fig1}.

In addition to the frequency dependence in Eq.~\eqref{EqHam87}, the numerical evaluation of the DoS shows a strong suppression of $\delta\nu_{\kappa}(\omega)$ for $\kappa>1$ with a factor of $\sim10^{-7(\kappa-1)}$. Apart from this strong numerical suppression, Eq.~\eqref{EqHam87} shows that for small $\omega$ the correction $\delta\nu_{\kappa}(\omega)$ has for $\kappa>1$ a subleading scaling compared to the bare DoS $\nu_0(\omega)\sim \omega^2$. We will focus on the leading order, $\kappa=1$ correction in the following and show that it does not modify the paradigm of a vanishing zero-energy DoS for Weyl semimetals. Due to their strongly subleading nature, this conclusion will hold for any $\kappa>1$ equally well. 

\begin{figure}
	\includegraphics[width=\linewidth]{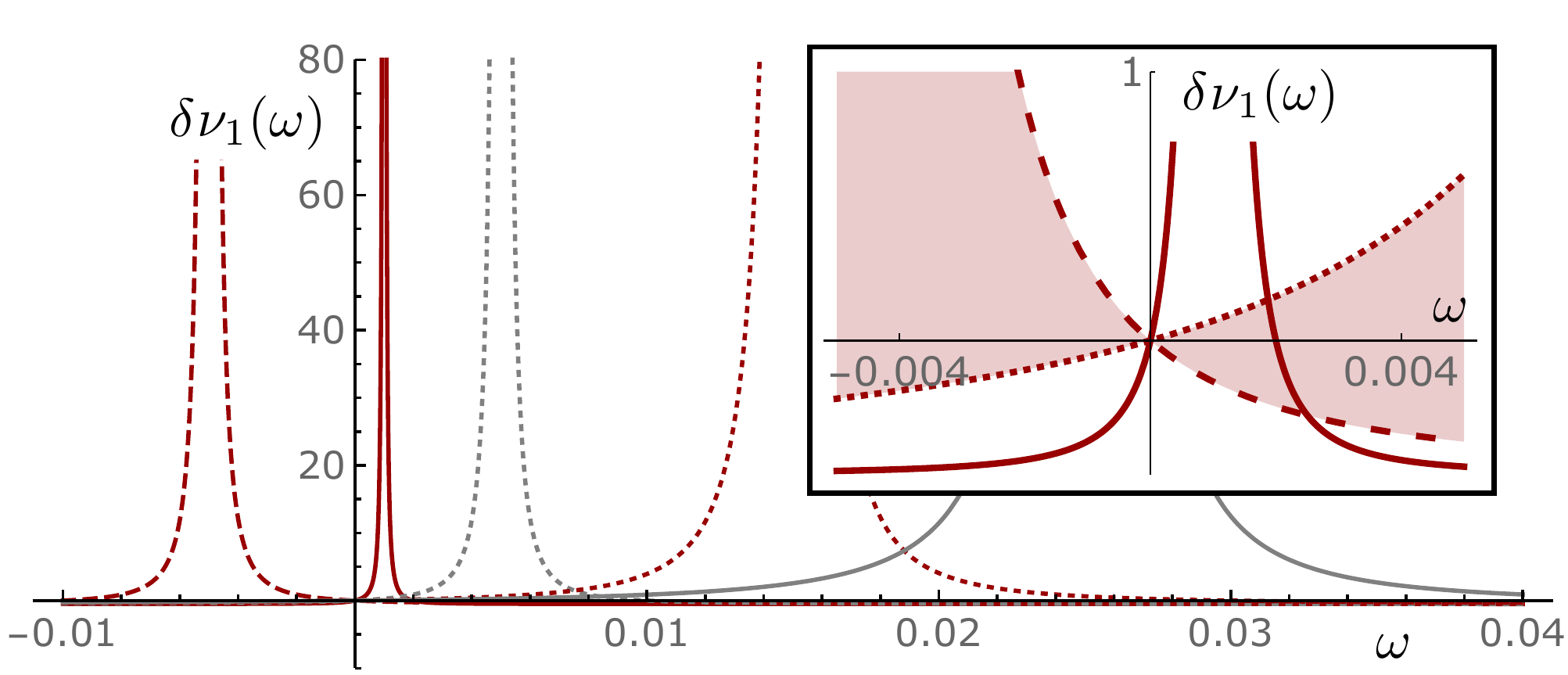}
	\caption{Density of states, $\delta\nu_{\kappa}(\omega)$ for $\kappa=1$ for a deviation $\Delta_{1}$ that approaches and passes zero (central figure from right to left $\Delta=(0.05, 0.03, 0.01, 0.002, -0.01)$). Inset: Focus on the vicinity of $\omega=0$ for the red colored configurations (right to left $\Delta=(0.03, 0.002, -0.01)$). While fulfilling $\delta\nu_1(\omega=0)=0$ for all configurations, the resonance peak becomes more and more narrow for $\Delta\rightarrow0$ and the accumulated spectral weight gets confined closer and closer to zero.
	}
	\label{fig2}
\end{figure}

In the following, we focus on the $\kappa=1$ sector and drop the angular momentum
label in the DoS and the potential, i.e.  we set
$\delta\nu(\omega)\equiv\delta\nu_{\kappa=1}(\omega)$,
$\Delta\equiv\Delta_{\kappa=1}$. As we have discussed, the DoS
$\delta\nu(\omega=0)=0$ is always zero. The frequency region in which a resonant
state for $\Delta\neq0$ accumulates spectral weight is, however, confined to the
region between $\omega=0$ and $\omega=\Delta$, see Fig.~\ref{fig2}. This region
becomes more and more narrow the closer $\Delta\rightarrow0$ approaches zero. On the
other hand, the resonance peak in $\delta\nu(\omega)$ also moves to zero and thus
spectral weight may be shifted closer and closer to $\omega=0$.

For a statistical distribution of spherical box potentials, the single impurity counterpart of a disordered system, the density of states at $\omega=0$ is guaranteed to be zero according to the discussion above. It is, however, by no means guaranteed that the average density of states remains continuous in the vicinity of $\omega=0$. Its continuity behavior depends on the limiting behavior of $\nu(\omega)$ for $\Delta\rightarrow0$ and will be investigated in the next section.

\subsection{Statistical distribution of the DoS} 
\label{sub:statistical_distribution_of_the_dos}

In view of the above findings, it seems a good idea to describe the DoS in terms of its probability distribution over an ensemble of box parameters. To this end, assume  the shifts $\Delta=\lambda-\pi$  distributed according to a probability distribution $P(\Delta)$ symmetric around zero and of width $\Delta_0\ll \pi$ (we aim to characterize the statistics of the DoS in near resonant cases), e.g. 
\begin{align}
\label{EqHam89}  
P(\Delta)= \frac{1}{\Delta_0\sqrt{\pi}}\exp{\left(-\frac{\Delta^2}{\Delta_0^2}\right)}.
\end{align} 
Each value of $\Delta$ generates a  DoS $\delta\nu(\omega,\Delta)$ as a dependent random variable.  The corresponding unit normalized distribution for the DoS is obtained via
\begin{align}
\label{EqHam88}
P(\rho,\omega)=\int_\Delta \delta\left(\rho-\delta\nu(\omega,\Delta)\right)P(\Delta).    
\end{align}
For $|\omega|, |\Delta|\ll \pi$, the phase shift is well approximated by
\begin{align}
\label{EqHam90}
\delta_{1}(\omega)\overset{|\omega|, |\Delta|\ll 1}{=}\arctan \frac{\omega^2}{2\omega-\Delta}.    
\end{align}
and the corresponding DoS $\delta\nu(\omega)=-\frac{1}{\pi}\partial_\omega \delta_1(\omega)$ leads to two distinct zeros $\Delta_\pm$ for the argument of the $\delta$-function ${\rho-\delta\nu(\omega,\Delta_\pm)=0}$. This yields the distribution
\begin{align}
\label{EqHam91}
P(\rho,\omega)&=\sum_{\alpha=\pm} \frac{P(\Delta_{\alpha})}{|\partial_\Delta \delta\nu(\omega,\Delta_\alpha)|},\\
\text{with }\Delta_{\pm}&=|\omega|\left(2+\frac{1\pm\sqrt{-\pi ^2 \rho ^2 \omega ^2+2 \pi  \rho  +1}}{\pi  \rho }\right).\nonumber    
\end{align}
The zeros under the square root determine the support of $P(\rho,\omega)$, i.e. it is nonzero only for $\rho\in[\rho_-,\rho_+]$ with ${\rho_{\pm}=\frac{1\pm\sqrt{1+\omega^2}}{\pi\omega^2}}$.

\begin{figure}
    \includegraphics[width=\linewidth]{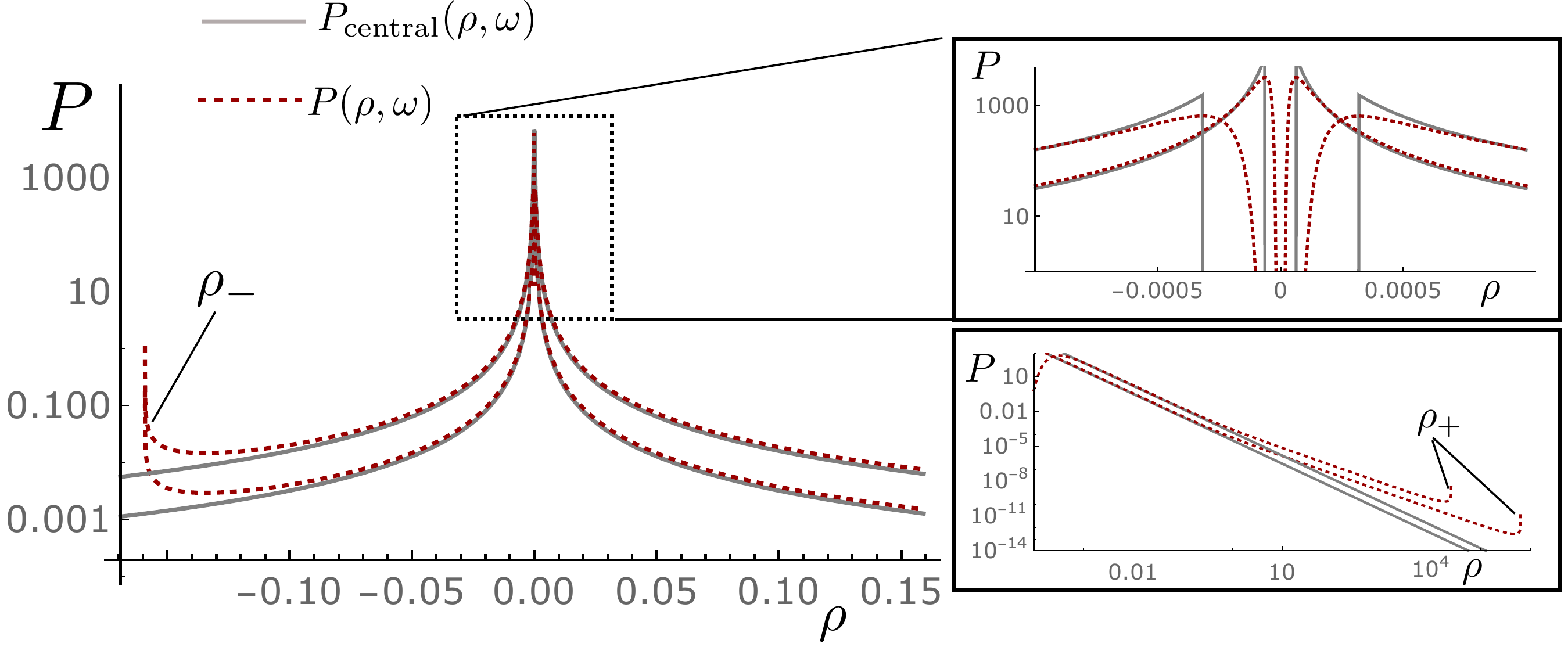}
    \caption{Probability distribution $P(\rho,\omega)$ of the DoS $\delta\nu(\omega)=\rho$ for fixed width $\Delta_0=0.1$ and different $\omega=10^{-5}, 5\cdot 10^{-5}$ (upper, lower lines) on a logarithmic scale. Left: Comparison of the distribution $P(\rho,\omega)$ (red, dashed) with the approximation for the central peak Eq.~\eqref{EqHam92} (gray, solid). Top right: Zoom into the region $\rho\approx 0$. Bottom right: Large $\rho$ behavior including the peak at $\rho=\rho_+$. 
    }
    \label{fig3}
\end{figure}

For fixed $\omega\neq0$, $P(\rho,\omega)$ has three different significant contributions, see Fig.~\ref{fig3}: i) a peak around $\rho=0$, which is symmetric in $\rho\rightarrow-\rho$ and expresses the equal probability for the DoS to be slightly raised or lowered by a nearby resonance, or to remain unmodified by the impurity, ii) a diverging peak at $\rho=\rho_-$, reflecting the lowest value the DoS can obtain at a given frequency by pulling spectral weight away and towards a resonance, iii) a diverging peak at $\rho=\rho_+$ which corresponds to the maximum spectral weight that can be acquired by a single resonance.  

For $\omega\rightarrow0$, one finds $\rho_-\rightarrow -\frac{1}{2\pi}$ and $\rho_+=\frac{1}{\omega}\rightarrow\infty$.
Recalling the behavior of $\delta\nu(\omega)$ from Fig.~\ref{fig2} (see also the inset) both the negative peak $\rho\sim-\frac{1}{2\pi}$ as well as the positive peak $\rho\sim \frac{1}{\omega}$ correspond to resonances that are very close to $\omega=0$. On the other hand, the peak centered around $\rho=0$ corresponds to resonances that have some minimal value $|\omega|\ge|\Delta|$. 

Analyzing these contributions separately, one finds that, as the energy approaches zero, the central peak acquires more an more statistical weight and, in the limit $\omega\rightarrow0$ its statistical measure approaches unity. This reflects the vanishing measure for hitting the exact value $\Delta=0$ in a given impurity realization. Consequently, the measure of the two peaks at $\rho=\rho_{\pm}$ vanishes in the limit $\omega\rightarrow0$. 

Let us explore what consequences the above structures have for the distribution of the DoS $P(\rho,\omega\simeq 0)$ at zero energy. Any DoS near zero energy is supported by $\Delta$ likewise near zero, a regime where the distribution $P(\Delta)$ is approximately constant. Integration of the derivative of Eq.~\eqref{EqHam90} over a range of $\Delta$ leads to the finite result $\langle \rho\rangle=\langle \delta\nu(0)\rangle_\Delta\sim \Delta_0^{-1}$. How can this finding be reconciled with a vanishing DoS at zero energy for each realization? The resolution is that this is the case of a singular distribution of the DoS, reflecting the fact that the average of the DoS over a continuous distribution of parameters $\Delta$ has nothing to do with the DoS observed in a single sample. To see how this happens, note that for $\omega\to 0$, the above peak in the distribution $P(\rho,\omega)$ disappears to $\rho\to \infty$ (peaks in the DoS asymptotically close to zero become infinitely large) and at the same time loses in height (they become increasingly rare). In the limit, these peaks represent events of probability measure zero: with probability unity the events they describe will not be seen in any single sampling from the $P(\Delta)$ distribution To illustrate this by an analogy, consider the `random variable' $X\equiv \delta(x-a)$ which equals $\infty$ if the continuous variable $x=a$ and zero else. For any sampling of $x, X$ equals zero. At the same time, the average over a continuous $x$-distribution, $\langle X\rangle_x=1$ and nonzero. 

The statistical description can be substantiated somewhat by exploiting that in the limit $\omega\to 0$, the full weight of the distribution is concentrated around $\rho=0$. Inspection of the distribution  shows that its central peak is well approximated by (cf. Fig.~\ref{fig3})
\begin{align}
\label{EqHam92}
P_\mathrm{central}(\rho,\omega)\equiv \frac{|\omega|}{\pi\Delta_0 \rho^2}\theta\left(|\rho|-\frac{2|\omega|}{\pi\Delta_0}\right)1_{[\rho_-,\rho_+]}(\omega),    
\end{align}
where $1_{[a,b]}$ is the characteristic function on an interval $[a,b]$. This distribution is unit-normalized up to corrections of $\mathcal{O}(\omega)$, and in the limit $\omega\to 0$ carries the full distribution. Computing the average DoS over $P_{\mathrm{central}}$, we find
\begin{align}
\label{EqHam95}  
\langle \delta\nu(\omega,\Delta)\rangle=\frac{|\omega|}{\pi\Delta_0}\log\left|\frac{\rho_+}{\rho_-}\right|\simeq \frac{|\omega|}{\pi\Delta_0}\log|\omega|.  
\end{align}
This demonstrates that in a description where unphysical rare events are filtered out, the average DoS vanishes linearly, on a broad scale set by the width of the parameter distributions.

This analysis shows that a single, spherical impurity is unable to generate a DoS at frequency $\omega=0$ for a Weyl particle and demonstrates that the average DoS for a statistical distribution of impurities vanishes continuously $\sim |\omega|\log|\omega|$ as $\omega\rightarrow0$ is approached. This result supports our statement that weak disorder is unable to generate a DoS at zero frequency from the perspective of a randomly distributed, single impurity.

{\color{black}Before concluding this section, let us briefly contrast the behavior discussed above to that of a Dirac metal Hamiltonian. The latter is equivalent to the superposition of two Weyl nodes at the same point in the Brillouin zone. Any type of scattering will now couple the two Weyl sectors and this leads to radically different structures in the bound state spectrum. Specifically, such systems support an infinite set of bound states for a continuous range of box potentials \cite{Goswami2016,Roy2017}. This results in a finite measure for rare region effects and renders the zero energy DoS of the Dirac system finite.}

\section{Vanishing density of states for a Gaussian disorder potential}\label{SUSY}
The vanishing of $\nu(0)$ in the previous section was obtained for the specific case of a box potential of exceptionally high symmetry. Since the result is not protected by general criteria, one may worry that it is an artifact of the model. It is therefore important to extend the analysis to more general, and randomly distributed potential profiles. This is the subject to which we turn next.

In this section, we will replace the box potential by a Gaussian distributed random potential with zero mean $\langle V_x\rangle_V=0$ and  correlation function 
\begin{align}\label{EqHam2}
\langle V_x
V_{y}\rangle_V =W^2\exp(-|x-y|/\xi).
\end{align}
Here $\langle ... \rangle_V$ is the average over a distribution characterized by the characteristic potential strength $W$, and correlation length $\xi$. (Broadly speaking, $W$ and $\xi$, now replace the parameters $\lambda$ and $r_0$ of the previous section.)

The analysis below is based on supersymmetric field integration methods. Our problem requires the careful analysis of fluctuation integrals and hence will be somewhat technical. The main result will be the identification of an  upper bound for $\nu(0)$, double exponential as $\sim \exp(- c\exp(c' (v/\xi W)^2))$ in the small disorder concentration ($c,c'$ are numerical constants). The small uncertainty has to do with the neglect of correlations between different rare event fluctuations in the analysis below. Readers ready to accept the statement that isolated strong potential flucutations (of box type as in the previous section, or of the more realistic Gaussian distributed form as below) do not generate zero energy DoS are invited to proceed to the next section~\ref{MultImp} where this gap is filled and 
rare event correlations are addressed.  

\subsection{Hamiltonian and supersymmetric action formalism}

Our main observable of interest in this section is the disorder averaged, retarded (advanced) Green's function at frequency $\omega$
\begin{align}\label{EqHam3}
G^{\pm}_{\omega,x,x'}=\left\langle \langle x | \frac{1}{\omega^\pm-\hat{H}}| x'\rangle\right\rangle_V,
\end{align}
where $\omega^{\pm}=\omega\pm i0^+$ and $x,x'$ are position arguments. The elements $G^\pm_{\omega,x,x'}$ are $2\times2$ matrices in Weyl spinor space from which the DoS is obtained as
\begin{align}\label{EqHam5}
\nu(\omega)=-\frac{1}{2\pi L^3} \int_x\text{Im} \ \text{tr}(G^{+}_{\omega,x,x}).
\end{align}
Here, $\int_x\equiv \int d^3x$ abbreviates the integral over three-dimensional space, $L$ is the linear dimension of the system and tr$(...)$ is the trace in Weyl spinor space. In the clean case, $V_x=0$, a straightforward evaluation of Eq.~\eqref{EqHam5} in momentum space and yields the quadratic DoS of the Weyl Hamiltonian $\nu(\omega)\sim \frac{\omega^2}{\nu_0^3}$. 

In the following, we review how the disorder average of the Green's function is
performed by supersymmetric path integral techniques (for a general introduction to
the approach, see Ref.\cite{EfetovBook} and for a previous application to the Weyl fermion
case Ref.~\cite{GurarieWeyl}). We start by introducing the superfields
$\psi_x=(\phi_x,\chi_x)^T$ and $\bar{\psi}_x=(\bar{\phi}_x, \bar{\chi}_x)$, where
$\bar{\phi}_x, \phi_x$ are spinors of commuting (complex valued) components and
$\bar\chi_x, \chi_x$ are spinors of anti-commuting (Grassmann) components. Next define
the supersymmetric action
\begin{align}\label{EqHam6}
S_V[\psi]=\int_x \bar\psi_x \left(\begin{array}{cc}\omega^+-\hat H&0\\0 &\omega^+-\hat H\end{array}\right)\psi_x.
\end{align}
 Application of standard rules for Grassmann and complex Gaussian integrals shows that 
\begin{align}\label{EqHam7}
G^{+}_{\omega,x,x'}=\left\langle\int \mathcal{D}[\bar\psi, \psi]\ (\bar\psi_{x'}\tau_3\psi_{x}) e^{iS_V[\psi]}\right\rangle_V,
\end{align}
with the Pauli matrix in superspace $\tau_3=\text{diag}(1,1,-1,-1)$, the integration measure $\mathcal{D}[\bar\psi, \psi]=\mathcal{D}[\bar\phi,\phi]\mathcal{D}[\bar\chi, \chi]$.
Rewriting the integral as
\begin{align}
    \label{EqHam9}
    G^{+}_{\omega,x,x'}=\int \mathcal{D}[\bar\psi, \psi]\ (\bar\psi_{x'}\tau_3\psi_{x})e^{iS_{\text{eff}}[\psi]},
\end{align}
where $e^{iS_{\text{eff}}[\psi]}\equiv \left\langle e^{iS_V[\psi]}\right\rangle_V$, we perform the Gaussian average to obtain the effective action as
\begin{align}\label{EqHam11}
     S_{\text{eff}}[\psi]=\int_x \,\bar \psi_x\Big(\omega^++iv_0\sigma_i\partial_i+iW^2\int_{y}e^{-\xi^{-1}|x-y|}\bar\psi_{y}\psi_{y}\Big)\psi_x.\end{align}
Like $S_V$, this action still exhibits supersymmetry and remains invariant under a uniform rotation of the superfield $\psi$. 

The expressions for the Green's function $G^+_{\omega,x,x'}$, Eq.~\eqref{EqHam9} and
the disorder averaged action $S_{\text{eff}}$, Eq.~\eqref{EqHam11}  are the  starting
point for the field integral analysis of the DoS for the disordered Weyl problem.
Following previous work (building on the closely related replica
formulation)\cite{Fradkin86a, GurarieWeyl3} one way to proceed would be to apply
renormalized perturbation theory   to investigate the long wavelength physics of the
problem. This leads to the prediction of quantum criticality mentioned in the
beginning. We here follow the alternative route~\cite{Nandkishore:2014aa} to identify
spatially localized field configurations (`instantons') $\psi_I$  extremizing  the action $S_{\text{eff}}$.
Although statistically rare, such configurations considerably modify the Green's
function \cite{Lutt,Rossum1994,Langer66}. It is natural to expect that in the vicinity of the nodal point,
$\omega\approx 0$, where the DoS of the clean Weyl system vanishes $\sim \omega^2$,
the presence of  instanton solutions qualitatively changes the behavior of   $G^+_{\omega,x,x'}$
and $\nu(\omega)$.

\subsection{Instanton calculus for the Green's function}
Instanton configurations $\psi_I$ are spatially localized solutions of the theory's
mean field equation\cite{Coleman78,Langer66, Langer67, Yaida2016}. Reflecting the
unstable nature of the resonances they represent, the instantons lie outside the
integration contour of $\psi$-variables.
To identify them, we first apply the transformation $(\psi,\bar
\psi)\rightarrow e^{ i\pi/4}(\psi_I^{\phantom{\dagger}},\bar \psi_I)$~\cite{Yaida2016,Lutt,Nieu}. In the new variables, the variational equations assume the form of a non-linear Schr\"odinger equation 
\begin{align}\label{EqHam12}
      \frac{\delta S}{\delta\bar\psi_I}=\left[\omega+iv_0 \sigma_i \partial_i-2W^2\int_y e^{-\xi^{-1}|x-y|}\bar\psi_{I,y}\psi_{I,y}\right]\psi_{I,x}^{\phantom{\dagger}}=0.\ \ \ \ \ \ \ \ \end{align}
Note that the operator in brackets, as well as the action~\eqref{EqHam11} from which it is derived is supersymmetric in that it remains invariant under spatially uniform rotations, $\psi\to U \psi$, where $U$ couples commuting and anti-commuting variables but is proportional to unity in Weyl space (see Eq.~\eqref{EqHam30} for an explicit representation of these transformations).    Without loss of generality, we can thus assume a solution  $\psi_{I,x}=(\phi_{I,x},0)^T$  living entirely in the commuting sector. Inserting this ansatz and taking the limit $\omega\rightarrow 0$ the saddle point equation reduces to
\begin{align}\label{EqHam13}
\left[iv_0 \sigma_i \partial_i-2W^2\int_y e^{-\xi^{-1}|x-y|}|\phi_{I,y}|^2\right]\phi_{I,x}^{\phantom{\dagger}}=0.
\end{align}
Neglecting the effect of fluctuations around the instanton solution, the Green's function is obtained by inserting solutions of Eq.~\eqref{EqHam13} into Eq.~\eqref{EqHam9},
\begin{align}\label{EqHam14}
G^{+}_{\omega,x,x'}&=-i \phi_{I,x}\bar\phi_{I,x'} e^{-S_I},\\
S_I&=W^2\int_{x,x'} |\phi_{I,x}|^2e^{-\xi^{-1}|x-x'|} |\phi_{I,x'}|^2\sim W^2\xi^2/v^2.\nonumber
\end{align}
This result was first obtained in Ref.~\cite{Nandkishore:2014aa}. It suggests a non-vanishing density of states at zero energy 
\begin{align}\label{EqHam15}
\nu(0)&=-\frac{1}{2\pi L^3}\int_x \text{tr}\ G^{+}_{0,x,x}=\frac{||\phi_I||^2e^{-S_I}}{2\pi L^3}.
\end{align}

Analytical solutions to Eq.~\eqref{EqHam13} can be found under the self-consistent approximation that the solutions themselves vary smoothly on scales $x\sim\xi$ \cite{Nandkishore:2014aa}. These solutions, which we will discuss in more detail below,  lead to a saddle point DoS\cite{Nandkishore:2014aa}
\begin{align}\label{EqHam16}
\nu(0)\sim \exp\left(-C\frac{\nu^2}{W^2\xi^2}\right)
\end{align}
with a numerical prefactor $C=O(1)$. This result suggests a non-vanishing $\nu(0)$ for arbitrarily weak disorder, non-perturbative in $W$ and $\xi$. 

In the following, we demonstrate that the above saddle point estimate does not survive the inclusion of fluctuations. Specifically, we will identify an extensive number of zero-action fluctuations in superspace. Broadly speaking, the presence of these fluctuations reflects the extreme instability of resonant states near zero energy, and their statistical insignificance as exemplified in the previous section for the box potential. Within the present formalism, integration over the zero modes suppresses the DoS down to zero, $\nu(0)=0$.

\subsubsection{Asymptotic instanton wave function}
Before turning to the discussion of fluctuations, we need to understand the asymptotic behavior of the solutions of Eq.~\eqref{EqHam13} at length scales $r\gg \xi$ and $r\ll\xi$, respectively. Assuming a solution centered around $x_0=0$ and denoting $\varphi_I(r,\theta,\phi)$ as the instanton wave function in spherical coordinates, we express spatial coordinates in units of the disorder correlation length and energies in terms of the effective disorder strength $\gamma=\tfrac{W^2\xi}{2\pi\nu_0}$. 

A detailed review of the solution for $r\gg1$ can  be found in the literature~\cite{Nandkishore:2014aa}. Under the self-consistent approximation of smooth solution behavior for  $r\gg\xi$ the instanton wave function has smooth variation, the integral kernel in Eq.~\eqref{EqHam13} can be replaced by a $\delta$-function, $ \exp(-|x-y|/\xi)\sim \xi^3 \delta^{(3)}(x-y)$, which leads to the local equation
\begin{align}\label{EqHam17}
\left[\frac{i}{r}\vec{\sigma}\cdot \vec{r}\left(\partial_r-\frac{\vec{\sigma}\cdot\vec{L}}{r}\right)-4\pi \gamma|\varphi_{I}|^2\right]\varphi_{I}^{\phantom{\dagger}}=0.\ \
\end{align}
Here, $\vec{L}=-i\vec{r}\times\nabla$ is the orbital angular momentum satisfying the  commutation relations
\begin{align}\label{EqHam18}
\left[\vec{J},\vec{\sigma}\cdot\vec{r}\right]=\left[ \vec{J},\vec{\sigma}\cdot\vec{L}\right]=0,
\end{align}
where $\vec{J}=\vec{L}+\frac{1}{2}\vec{\sigma}$ is the total angular momentum.
The operator $O_{\vec{r}}=\frac{1}{r}\vec{\sigma}\cdot\vec{r}$ does not commute with the orbital angular momentum and exchanges states with different parity 
\begin{align}\label{EqHam19}
\left(O_{\vec{r}}\right)^2=\mathds{1}, \ \ O_{-\vec{r}}=-O_{\vec{r}}, \ \ \left[\vec{L}^2,O_{\vec{r}}\right]\neq0.
\end{align}
Solutions of Eq.~\eqref{EqHam17} can thus be defined in subspaces of conserved half integer total angular momentum $j$, $\vec{J}^2=(j+1)j$, shared between  spin $s=\pm \frac{1}{2}$ and  orbital angular momentum $l=j\mp\frac{1}{2}$. Introducing $|\uparrow,\downarrow\rangle$ for the spin states in Weyl space, the instanton wave functions can be expanded in spherical harmonics as
\begin{align}\label{EqHam20}
    &\varphi_I(r,\theta,\phi)=\sum_{l=0,1}y_l(\theta,\phi) \,i^l\, f_l(r),\\
    &\quad y_0=|\uparrow\rangle \otimes Y_{00},\quad y_1=|\uparrow \rangle \otimes Y_{1,0}-\sqrt{2}|\downarrow \rangle \otimes Y_{1,1}.\nonumber
\end{align}
Insertion of this ansatz into Eq.~\eqref{EqHam17} leads to the two coupled equations 
\begin{align}\label{EqHam21}
\gamma\left(f_1^2+f_0^2\right)f_m=i^{2l}(\delta_{l,m}-1)\left(\partial_r+\frac{2\delta_{l,1}}{r}\right)f_l,
\end{align}
which are solved by the series
\begin{align}\label{EqHam22}
f_l(r)=\frac{1}{\sqrt{\gamma}r^2}\sum_{k=0}\eta_{l,k}
\,r^{-6k-3(1-l)}.
\end{align}
The coefficients $\eta_{l,k}$ are dimensionless and real, and for $k\rightarrow\infty$ rapidly approach zero. One observes that $f_0$ decays much faster in $r$ than $f_1$, such that on the largest distances $f_{0}\sim 0$ and $f_1\sim 1/(\sqrt{\gamma}r^2)$. The approximation of smoothly varying solutions is fulfilled by Eq.~\eqref{EqHam21} a posteriori. 

On short distances, $r\ll1$,  continuity of the wave function for $r\rightarrow0$ requires that the vanishing of the finite angular momentum component, $f_1(r\rightarrow0)\rightarrow 0$,  and constancy of the zero component $f_0(r\rightarrow0)\rightarrow $const. 

Although these  criteria are sufficient for the fluctuation analysis below, we will briefly discuss an approximate solution for $r\ll1$: The assumed smoothness of the solution allows us to approximate Eq.~\eqref{EqHam13} as
\begin{align}\label{EqHam23}
\left[\frac{i}{r}\vec{\sigma}\cdot \vec{r}\left(\partial_r-\frac{\vec{\sigma}\cdot\vec{L}}{r}\right)-4\pi \gamma m\right]\varphi_{I}^{\phantom{\dagger}}=0,
\end{align}
where $m=\int_V \exp(-r)|\varphi_I(r)|^2$ acts as an effective energy shift. The  solutions of this equation are mathematically identical to those of the scattering wave functions in a box potential,
\begin{align}\label{EqHam24}
\varphi_I(r,\theta,\phi)=y_0(\theta,\phi)\left(A_JJ_{j+\frac{1}{2}}(\gamma mr)+A_Y Y_{j+\frac{1}{2}}(\gamma mr)\right),
\end{align}
where $y_0$ is defined in Eq.~\eqref{EqHam20}, $A_J, A_Y$ are numerical prefactors and $J, Y$ are the spherical Bessel functions of the first and second kind.

\subsubsection{Global zero modes}
Turning to fluctuations around $\psi_I$, we first consider uniform fluctuactions, constant in space. Any variation of $\psi_{I,x}$ can be decomposed into  unitary rotations of the instanton wave function in superspace $\psi_x\rightarrow U\psi_{I,x}$, and amplitude fluctuations of the instanton wave function $\psi_x\rightarrow \psi_{I,x}+\delta\psi_I$. This allows us to parametrize the field as $\psi_x=U(\psi_{I,x}+\delta\psi_I)$, where $U$ and $\delta\psi_I$ do not depend on $x$. Inserting this into the action \eqref{EqHam11} and exploiting that $\partial_iU=\partial_i\delta\psi_I=0$, one finds 
\begin{align}\label{EqHam25}
S_{\text{eff}}[U(\psi_{I,x}+\delta\psi_I)]=S_{\text{eff}}[\psi_{I,x}+\delta\psi_I],
\end{align}
i.e. all variations parametrized by the unitary transformation $U$ have zero action. 
The inclusion of global fluctuations into the expression for the Green's function gives
\begin{align}\label{EqHam26}
G^{+}_{\omega,x,x'}&=\int \mathcal{D}[\bar U, U] \mathcal{D}[\delta\bar\psi, \delta\psi ]\Big[(\bar\psi_{I,x'}+\delta\bar\psi_I)\bar U\tau_3U(\psi_{I,x}+\delta\psi_I) \nonumber\\
& \times\ e^{iS_{\text{eff}}[\psi_{I,x}+\delta\psi_I]}\Big],
\end{align}
where the measure $\mathcal{D}[\bar U, U]$ will be discussed below. Since $U$ does not appear in the action, it contributes free integrations to the Green's function \eqref{EqHam9} (and DoS), which need to be treated with some care. 

The fluctuations encoded in $U$ are closely related to the continuous symmetries in the original action $S_{\text{eff}}$ that are explicitly broken by the instanton solution $\psi_{I,x}$. The symmetry of  $S_{\text{eff}}$ implies that $U\psi_{I,x}$ is also a solution of Eq.~\eqref{EqHam12}, continuously connected to $\psi_{I,x}$ by a path of zero action. 

The set of fields generated in this  way is embedded into  the Hilbert space $\mathcal{H}$ of  superfields $\psi_x\in\mathcal{H}$, 
\begin{align}\label{EqHam26}
\mathcal{H}=L^2(\Bbb{R}^3)\otimes
\Bbb{C}^2\otimes\Bbb{C}^{1|1},
\end{align}
where $L^2(\Bbb{R}^3)$ is the space of scalar, square-integrable wave
functions over $\Bbb{R}^3$, $\Bbb{C}^2$ the space of Weyl-spinors with two complex components, and $\Bbb{C}^{1|1}$ the two-dimensional space of superfields with anti-commuting and
commuting components. The action $S_{\text{eff}}$ is invariant under translations in $\Bbb{R}^3$, generated by the three momentum operators $p_l=-i\partial_l$, rotations in $L^2(\Bbb{R}^3)\otimes
\Bbb{C}^2$ generated by the three total angular momentum operators $J_i$, and under supersymmetric rotations in $\Bbb{C}^{1|1}$. These are seven continuous symmetries, each  broken by $\psi_{I,x}$.

Individual broken symmetry transformations of commuting degrees of freedom define  bosonic zero modes\cite{Lutt,Rossum1994,Langer66,Yaida2016}. In the present context, these modes are the three spatial coordinates of the instanton's origin $\vec{r}_0=(x_0,y_0,z_0)$ and the three rotation angles $\vec{\theta}=(\theta_1,\theta_2,\theta_3)$. The transformations generated by these modes can be represented as 
\begin{align}\label{EqHam27}
U_{L^2(\Bbb{R}^3)\otimes
\Bbb{C}^2}=\exp(r_i\partial_i)\cdot \exp(i\theta_iJ_i).
\end{align}
These fluctuations have zero action \eqref{EqHam25} and do not appear in the prefactor $\bar\psi_{x}\tau_3\psi_x$ in Eq.~\eqref{EqHam9} for the Green's function at equal coordinates $x=x'$. The mode integration thus leads to a global prefactor $\int d^3rd^3\theta=8\pi^3L^3$ multiplying the local Green's function. Specifically,  the volume prefactor  is  required for the normalization of the instanton DoS \eqref{EqHam15}  in the thermodynamic limit  $L\rightarrow\infty$\cite{Lutt,Rossum1994,Langer66,Yaida2016,Halperin66,Nieu}. 

Super-rotations of the commuting configuration  $\psi_{I}=(\varphi_I,0)^T$ can be parameterized as\cite{EfetovBook}
\begin{align}\label{EqHam30}
\exp\left(\begin{array}{cc}0& \Xi\\ \bar\Xi & 0\end{array}\right)=\left(\begin{array}{cc} \mathds{1}+\frac{1}{2}\Xi\bar\Xi& \Xi\\ \bar\Xi & \mathds{1}+\frac{1}{2}\bar\Xi\Xi
\end{array}\right),
\end{align}
where $\Xi, \bar\Xi$ are $2\times2$ matrices in Weyl spinor space. These are symmetries provided that the symmetry generators $\Xi$ commute with all the Pauli matrices, $\sigma_i$, $\Xi=\eta \mathds{1}$, with a single anticommuting generator $\eta$. While the action remains invariant under transformations of the form \eqref{EqHam31} the prefactor $\bar\psi_x\tau_3\psi_x$ transforms according to
\begin{align}\label{EqHam32}
\bar\psi_x\bar U\tau_3U\psi_x&=\bar\psi_x\bar U_{\Bbb{C}^{1|1}}\tau_3U_{\Bbb{C}^{1|1}}\psi_x=\bar\psi_x\tau_3\psi_x+\bar\psi_x\tau_\eta\psi_x,\nonumber\\
\tau_{\eta}&=2\left(\begin{array}{cc}\eta\bar\eta& \eta\\ -\bar\eta& \eta\bar\eta\end{array}\right)\otimes\mathds{1}.
\end{align}
Finally, the measure $\mathcal{D}[\bar U, U]\sim d\eta d\bar\eta$ is `flat' in that it does not depend on the variables $\bar\eta, \eta$.

Integrals over Grassmann variables are defined as
\begin{align}\label{EqHam33}
\int d\eta =0, \ \int d\eta\ \eta=1. 
\end{align}
Since the action \eqref{EqHam25} is also free of Grassmann variables, the non-vanishing of the fluctuation integral hinges on the pre-exponential dependence $\sim \eta\bar\eta$ of Eq.~\eqref{EqHam32} on the variables $\eta$. 

Summarizing, the full set of uniform symmetry transformations is  given by
\begin{align}\label{EqHam31}
U=U_{\Bbb{C}^{1|1}}\times U_{L^2(\Bbb{R}^3)\otimes
\Bbb{C}^2}, \text{ with } U_{\Bbb{C}^{1|1}}=\exp\left(\begin{array}{cc}0& \eta\mathds{1}\\ \bar\eta\mathds{1} & 0\end{array}\right).
\end{align}
While the detailed form of the integration measure $\mathcal{D}[\bar U, U]\sim d\eta
d\bar\eta$ depends on the shape of the instanton wave function $\psi_{I,x}$, the
discussion above contains an important message: Whether or not the saddle point
Green's function and DoS survive  may very well depend on the effect of fluctuations.
For any rotation in superspace, the prefactor in Eq.~\eqref{EqHam32} will contain
products of at most two different Grassmann variables $\eta, \bar\eta$ at the same
time. This implies that for any number $n>2$ of Grassmann zero modes the Grassmann
integral, and along with it the instanton contribution to the Green's function, will
vanish. As we will demonstrate below, this is precisely what happens for a disordered
Weyl semimetal.

\subsection{Instability of the saddle point and vanishing DoS}
The instanton approach as discussed thus far does not differ much from that of conventional disordered single particle problems. Both, the
non-perturbative dependence of the DoS on $W^2$ \eqref{EqHam15}, and the
contribution from global zero mode fluctuations are
common features\cite{Lutt,Rossum1994,Langer66,Yaida2016,Halperin66,Nieu}. What \emph{is} specific to the
Weyl system is the algebraic (rather than exponential) decay of the instanton wave function
$\psi_{I}$. This, in combination with the linearity of the Weyl operator, leads to the emergence of an
infinite set of Grassmann zero modes, beyond the trivial zero mode discussed previously. For the reasons indicated above, the presence of these  modes makes the fluctuation integral vanish and in this way protects the integrity of the spectral node.

\subsubsection{An extended set of Grassmann zero modes} 
We now extend the discussion of the previous section in that we include fluctuations  varying in space, 
\begin{align}\label{EqHam34}
\psi_x= U_x(\psi_{I,x}+\delta\psi_x)\equiv U_x \varphi_x.\end{align}
This representation differs from Eq.~\eqref{EqHam26} in that $U_x$ now only contains rotations in superspace and $\delta\psi_x$ variations of commuting variables. Since the instanton wave function $\psi_{I,x}$ is purely bosonic either,  $\varphi_x$ is a field of commuting variables. The representation above defines a complete coverage of the superspace of integration variables. We note that fluctuations in $\varphi_x$ include the   bosonic Goldstone modes reflecting rotational and translational symmetry breaking. While these fluctuations need not be small  all that matters is that they are bosonic and do not generate singular fluctuation determinants \footnote{This latter assertion follows from an explicit expansion in bosonic Goldstone modes. Zero action fluctuation modes which may arise at the level of a second order expansion are regularized by nonlinear higher order contributions to the action. The corresponding expansion is cumbersome and we will not discuss it in detail.}.

Rotations $U_x$ in superspace are generated by Grassmann variables and we represent them as a product of two $4\times4$ matrices,
\begin{align}\label{EqHam35}
U_x&=W_x \cdot V_x\\
W_x&=\exp\left(\begin{array}{cc}0& \eta_x\mathds{1}\\ \bar{\eta}_x \mathds{1}&0\end{array}\right)=\left(\begin{array}{cc}(1+\frac{\eta_x\bar{\eta}_x}{2})\mathds{1}& \eta_x\mathds{1}\\ \bar{\eta}_x\mathds{1}&(1-\frac{\eta_x\bar{\eta}_x}{2})\mathds{1} \end{array}\right)\label{eqW},\\
V_x&=\exp\left(\begin{array}{cc}0& \Delta_x\sigma_z\\ \bar{\Delta}_x \sigma_z&0\end{array}\right)=\left(\begin{array}{cc}(1+\frac{\Delta_x\bar{\Delta}_x}{2})\mathds{1}& \Delta_x\sigma_z\\ \bar{\Delta}_x\sigma_z&(1-\frac{\Delta_x\bar{\Delta}_x}{2})\mathds{1} \end{array}\right),\ \ \ \ \ \ \
\end{align}
where $\mathds{1}$ and $\sigma_z$ act in  Weyl spinor space. 

This  parametrization factors into Goldstone modes $\eta_x, \bar\eta_x$, isotropic in
Weyl space and gapped non-Goldstone modes $\Delta_x, \bar\Delta_x$. (These modes, in
combination with those contained in the non-Grassmann fluctuation matrices span the
full Hilbert space, and the inclusion of Grassmann generators coupled to further
Pauli matrices would be an overcounting.) The $\Delta$-modes are gapped in that they
have finite action even if spatially constant, on account of their non-commutativity with the Weyl Hamiltonian. At the same time, their coupling to the
$\eta$-modes changes the action of the latter in a not quite innocent way. We
therefore avoid ignoring the $\Delta$-fluctuations from the outset but postpone their discussion to section~\ref{MassModes}. 

The change of variables $\psi_x\rightarrow U_x\varphi_x$ modifies the integration
measure of the path integral. In  appendix~\ref{IntMeas} we show  that the measure
changes as $d\psi_xd\bar\psi_x\rightarrow J_x\
d\varphi_xd\bar\varphi_xd\eta_xd\bar\eta_xd\Delta_xd\bar\Delta_x$ where $J_x=\prod_l
|\varphi_{x,l}|^2$ is the Jacobian and $l=1,2$ is the Weyl spinor index. 

Dropping the  spatial index $x$ for notational brevity, the action becomes  of the fields $\psi=U\varphi$ becomes 
\begin{align}\label{EqHam38}
    S[\eta,\Delta,\varphi]=v_0 \int_x \, \bar \varphi U^{-1}(\sigma_i\partial_i)U \varphi + S^{(4)}[\varphi].
\end{align}
Note that the quartic part $S^{(4)}$ is invariant under rotations and contains only
massive fluctuations $\varphi$. Denoting the part of the action that contains
Grassmann variables as $S[\eta,\Delta]$, an expansion of the rotation matrices yields
\begin{align}\label{EqHam39}
S[\eta,\Delta]&=\underbrace{\nu_0\int_x\bar\varphi \sigma_\alpha W^{-1}(\partial_\alpha W)\varphi}_{=S_g[\eta]}+\underbrace{\nu_0\int_x\bar\varphi V^{-1}[\sigma_\alpha\partial_\alpha,V]\varphi}_{=S_m[\Delta]}\nonumber\\
&+\underbrace{\nu_0\int_x\bar\varphi V^{-1}\sigma_\alpha [W^{-1}(\partial_\alpha W),V]\varphi}_{=S_{gm}[\eta,\Delta]}.
\end{align}
We focus on the Goldstone action $S_g[\eta]$ first and treat the remaining parts of the action in Sec.~\ref{MassModes}. 

The Goldstone action can be expanded as
\begin{align}\label{EqHam40}
S_g[\eta ]=v_0\int_x\, (\partial_i \bar \eta\  \eta -\bar
    \eta\ \partial_i \eta) \, \bar \varphi \sigma_i \varphi.
\end{align}
The nilpotency of the Grassmann fields make this action purely quadratic in
$\eta,\bar\eta$. The linearity in derivatives of the quadratic action is specific to the Weyl
system. In the following we demonstrate that the combination of these two features
implies the existence of an infinite set of zero modes.

A field fluctuation $\eta$ has vanishing action $S_g$ if it solves the differential equation
\begin{align}\label{EqHam41}
M^i\partial_i\eta=M^i\partial_i\bar\eta=0 \text{ with } M^i=\bar\varphi\sigma_i\varphi.
\end{align}
This is a first order linear partial differential equation which, absent singularities in $M^i$, has  an infinite number of solutions. Before discussing the construction of such solutions, let us address the consequences of their presence.  The anti-hermitian operator $M^i\partial_i$ has a set of eigenfunctions $F_{a,x}$, which vary in space and which we label with $a\in\mathbb{N}$. They solve the eigenvalue equation
\begin{align}\label{EqHam42}
M^i\partial_iF_{a}=\lambda_a F_a.
\end{align}
The anti-hermiticity of $M^i$ implies  $\lambda_a^*=-\lambda_a$ and the completeness of the function set $\{F_{a}\}$ in the Hilbert space of square integrable functions. Let us assume the existence of a large but finite number $n_g>0$ of eigenvalues $\lambda_a=0$ (included in this set is the previously discussed global zero mode $a=0$ with its constant $F_0$). We may then expand the Grassmann fields as
\begin{align}\label{EqHam43}
\eta=\sum_a F_a\eta_a,
\end{align}
in an infinite and discrete set of Grassmann variables $\eta_a, \bar\eta_a$.
Assuming orthonormalization of the $F_a$'s the action assumes the form
\begin{align}\label{EqHam44}
S_g[\eta]=2\nu_0\sum_{a>n_g} \lambda_a \bar\eta_a\eta_a,
\end{align}
and is independent of the  Grassmann zero modes. 

The prefactor in the path integral for the Green's function \eqref{EqHam9} can likewise be expanded and becomes
\begin{align}\label{EqHam45}
\bar \psi \tau_3 \psi=2\bar \varphi \varphi \eta \bar \eta=2\bar \varphi
\varphi \sum_{ab}\bar F_a F_b \eta_a \bar\eta_b.
\end{align}
Note that the maximum power of pre-exponential Grassmann variables equals two. 
For any $n_g>0$ this leaves a product of $2(n_g-1)$ uncompensated  zero mode integrals, $\int d\bar \eta_a =\int d\eta_b=0$. 

In other words, the existence of non-trivial zero modes implies the vanishing of the
DoS. In the physically different case of disordered Schr\"odinger operators, the
fluctuation operator acting on $\eta$ is a second order elliptic differential
operator with a gapped spectrum. In such cases, all modes besides $\eta_0$ have
finite eigenvalues, and the fluctuation integral yields a harmless fluctuation
determinant\cite{Lutt,Rossum1994,Langer66,Yaida2016,Halperin66,Nieu}. Within the
present formalism, this is the main difference between the two cases. 

The
mathematically exact vanishing of the  eigenvalues relies on the linearity in derivatives of the
fluctuation operator and is compromised in the physically realistic case of spectra
with finite band curvature, coupling to massive modes, or the imposing of large
volume real space cutoffs. Under these circumstances, the product of a large number
of zeros gets replaced by a large number of numbers, each parametrically small. As we will discuss belwo,  this
turns a zero fluctuation determinant into one double-exponential in the small
parameters of the theory, still vanishing from a physical point of view.

\subsubsection{Explicit construction of inhomogeneous zero modes} 
Before discussing the effects of  massive mode couplings and other perturbations of the strictly linear eigenfunction problem, we need to understand some essential characteristics of the zero mode eigenfunctions themselves. To this end, we consider a given instanton wave function $\varphi=\varphi_I$, expanded as in  Eq.~\eqref{EqHam20}. With the explicit form of the spherical harmonics  $
Y_{1,1}=-\sqrt{3/(8\pi)}\sin\theta e^{i\phi},$ and $Y_{00}=\sqrt{1/(4\pi)}, Y_{10}=\sqrt{3/(4\pi)}\cos(\theta)
$, it is then straightforward to verify that the spherical coefficients $(M^r,M^\theta,M^\phi)^T$ of the cartesian vector  $(M^1,M^2,M^3)^T$,
\begin{align}\label{EqHam47}
M^i\partial_i=M^r\partial_r+\frac{M^\theta}{r}\partial_\theta+\frac{M^\phi}{r\sin \theta}\partial_\phi.
\end{align}
are defined by 
$M^r=(f_1^2+f_0^2)\cos\theta$, $M^\theta=(f_1^2-f_0^2) \sin\theta$, and
$M^\phi=2if_0f_1 \sin \theta$. This brings the zero mode equation into a separable form
\begin{align}\label{EqHam48}
    &\left((f_0^2+f_1^2)\cos\theta \partial_r + \frac{f_1^2-f_0^2}{r}\sin\theta\partial_\theta + \frac{2if_0f_1}{r} \partial_\phi\right) F_a =0,
\end{align}    
which can be solved by a strategy familiar from the treatment of the hydrogen problem: 

Due to  azimuthal  symmetry  the differential operator in Eq.~\eqref{EqHam48} is
independent of $\phi$. We thus consider solutions  of definite orbital momentum   $L_z$
with integer eigenvalue $m$,
$F_a(r,\theta,\phi)=\exp(im\phi)F_{a,m}(r,\theta)$. Even the $m=0$ subspace contains an infinite number of zero modes and we focus on this sector for simplicity. The $m=0$
equation is solved by the ansatz $ F_{a,0}(r,\theta)=F_{a,0}\left(\sin\theta
e^{Q(r)}\right)$, whose insertion into Eq.~\eqref{EqHam48} yields
\begin{align}\label{EqHam49}
\left((f_0^2+f_1^2)Q'(r)+\frac{f_1^2-f_0^2}{r}\right)\cos\theta\sin\theta F_{a,0}'=0. 
\end{align}
For given $f_1, f_0$ this is solved by
\begin{align}\label{EqHam50}
Q(r)=-\int_{\infty}^r \frac{d\rho}{\rho} \frac{f_1^2(\rho)-f_0^2(\rho)}{f_0^2(\rho)+f_1^2(\rho)}.
\end{align}
Inserting the asymptotic solutions for $f_{1,2}$ in the limit $r\rightarrow\infty$ \eqref{EqHam23} and $r\rightarrow0$ \eqref{EqHam25}, one finds the asymptotic behavior
\begin{align}\label{EqHam51}
Q(r)\sim \log(r)\times\left\{\begin{array}{cl} (-1) & \text{ for } r\rightarrow\infty\\
1& \text{ for } r\rightarrow0\end{array}\right. .
\end{align}
The equations leave the function $F_{a,0}(y)$, $y\ge 0$ undetermined, and the freedom to choose different differentiable and linearly independent functions $F_{a,0}$ reflects the existence of multiple zero modes. The choice $F_a=\mathrm{const}.$ defines the trivial zero modes, and the  limiting profiles of Eq.~\eqref{EqHam51} guarantee that non-constant choices of $F_{a,0}$ define further solutions, differentiable throughout the entire parameter range. In this way an infinite set of zero modes is constructed. In the next section we discuss what happens to the degenerate zero-spectrum of these modes when the idealizing assumptions underlying the above solution are relaxed. 

\subsubsection{Parametric smallness in the coupling to massive modes}\label{MassModes}

In this section, we explore what happens if the set of pristine zero modes gets coupled to the massive modes. We will find that this induces a term of second order in derivatives in the effective action of the former. This term adds to the  presence of higher order derivative operators contained in a realistic Weyl Hamiltonian. While such terms render the effective $\eta$-action finite, we will identify a small parameter $y_0\ll 1$ in which the total fluctuation contributions multiply to a double exponentially small net result. 

Specifically, we will demonstrate that the set of functions $F_a$  includes a subset with support on distance scales  $r>y_0^{-1}\gg 1$ and  action proportional to $y_0$. We will first identify these functions and then discuss their role in the fluctuation integral.
The action $S_m[\Delta]$ of the massive Grassmann modes in  Eq.~\eqref{EqHam38} is dominated by the non-vanishing commutators  $[\sigma_i,V_x]$, compared to which the derivatives acting on $V_x$ are small and will be neglected throughout. Under this assumption and setting $\nu_0=1$ for simplicity, we find 
\begin{align}\label{EqHam52}
S_m[\Delta]&=\int_x \,|\varphi|^2\Delta \mathcal{V} \bar\Delta,\nonumber\\
\mathcal{V}&=|\varphi|^{-2} \bar\varphi \sigma_z\sum_{\alpha}[\sigma_\alpha,\sigma_z]\partial_\alpha\varphi,\\
S_{gm}[\eta,\Delta]&= \int_x
 |\varphi|^2\left[\Delta\partial_z\bar \eta-\bar\Delta\partial_z\eta+\Delta\bar\Delta\bar M^z\left(\bar \eta\partial_z\eta-\eta\partial_z\bar \eta\right)\right].\nonumber
 \end{align}
This action is quadratic in the variable $\Delta$ which can hence be integrated out. Due to the nilpotency of the Grassmann variables, $(\partial_z\eta)^2=0, \eta\partial_z\eta=\frac{1}{2}\partial_z\eta^2=0$, the term linear in the derivatives $\sim M^z$ will not contribute after the integration and can  be discarded. The integration over $\Delta$ thus produces a second order  correction to the $\eta$-action  
\begin{align}\label{EqHam55}
\delta S_g[\eta]&=\int_x |\varphi|^2 \partial_z\bar \eta \mathcal{V}^{-1}\partial_z\eta\\
&=\sum_{a,b\le n_g}\bar \eta_a \eta_b\underbrace{\int_x |\varphi|^2 \partial_z\bar F_a \mathcal{V}^{-1}\partial_zF_b}_{\equiv I_{a,b}},\label{EqHam56}
\end{align}
 which in the second line we have expanded in the zero mode eigenfunctions of the unperturbed problem. (We here ignore the feedback of the massive modes into the action of modes $\eta_a$ of non-vanishing native action.) In Appendix \ref{AppIntkern} we show that the integral kernel can be represented as 
\begin{align}\label{EqHam61a}
     I_{a,b}\le-2\int_0^{1/2}dy y^3 \ln(y/2)\,\bar F_a'(y)F_b'(y).
\end{align}
Now consider a set
of linearly independent functions with support on a narrow strip $y\in[0,y_0]$. Using that $y=\sin(\theta) \exp(Q(r))$ with (Eq.~\eqref{EqHam51})) $\exp(Q(r))=1/r$ and $\exp(Q(r))=r$ for large and small arguments, $r$, respectively, we note that such functions have their support in a wide real space interval  $r\sim [y,y^{-1}]$. 
Equation \eqref{EqHam61a} shows that up to logarithmic corrections, the action of such
modes has an upper bound  $I_{a,b}\le \sup_{y\in [0,y_0]}|F'_b(y)F'_a(y)| \ y_0^4$. This implies the existence of a large set of  functions $\{F_a\}$ whose  action is small in powers of $y_0$. (As an example, consider plane waves $F_n\sim \exp(i n y_0^{-1}y)$ for which $|F_n'|\sim ny_0^{-1}$, and the kernel becomes $\sim n^2 y_0^2$.)

\subsection{Instanton Weyl density of states} 
\label{sub:instanton_weyl_density_of_states}

The above estimate for the fluctuation action depends on the number of derivatives in
the induced $\eta$-action and holds equally for a situation where a second order
derivative (representing band structure curvature away from the Weyl points) is part
of the definition of the Weyl operator. Either way, we find a class of low action
modes, as discussed above. Realistically, we should contain an extended system to
contain a \emph{density} of rare event fluctuations (and not just one, as in our
discussion above.) In the field integral approach, this corresponds to variational
solutions containing the linear superposition of various instanton wave functions.
Since the characteristic `fugacity' of an instanton is given by $\exp(-S_I)$, the
balance $R^3 \exp(-S_I)$ between phase volume and action leads to a characteristic
spacing $R\sim \exp(S_I/3)$. This scale limits the maximal extension of the
fluctuation modes discussed in previous sections. In particular, it sets a lower
bound $y_\mathrm{min} \sim R^{-1}\sim \exp(-S_I/3)$ for the effective control
parameter of the previous section. By order of magnitude, the modes of lowest action
cost thus have $S_n\equiv n^2 y_\mathrm{min}^2 $, $n=0,\dots,y_\mathrm{min}^{-1}$,
where we cut the spectrum at action costs of order unity, where generic fluctuations
enter the stage.

The fluctuation determinant multiplying the saddle point thus is of order
$\prod_{n=1}^{y_\mathrm{min}^{-1}} (n y_\mathrm{min})\sim\exp(-2
y_\mathrm{min}^{-1})\sim \exp(-2 \exp(S_I/3))$. Although the above estimate is rather
coarse, the essence of the argument is the existence of a large set of modes of
parametrically small action. This is a robust feature leading to a prefactor
\emph{double}-exponential in the small parameters of the theory, and hence
effectively zero for all practical purposes. 

We conclude that the nodal density of states generated by large fluctuations in a  Gaussian distributed random potential is vanishing. This statement relies on the peculiar features of the spectrum of fluctuation modes around the Weyl instanton. For finite energies, all these modes require a finite action, and the impurity DoS becomes finite.

\section{Exact density of states in a multi impurity model} 
\label{MultImp}
In our analysis above, the focus has been on the density of states of isolated potential wells of fixed (box) shape, section~\ref{SingleImp}, or of a statistically distributed form, section~\ref{SUSY}. In this section, we approach the problem from a different perspective and focus on \emph{correlations} between different rare event configurations. This is of relevance inasmuch as hybridization effects between isolated potential extrema may shift the position of density of states resonances, including into the neighborhood of zero energy. We aim to explore if our previous finding of vanishing density of states survives this mechanism.

Besides the presence of multiple impurities another new aspect of the present section is that band structure curvature is included in the analysis. This additional feature will in fact be required to remove unwanted UV singularities appearing in the unphysical case of an infinitely extended linear spectrum. We thus assume a spectrum $\sim |k|+M^{-1}k^2$ showing curvature beyond a large momentum scale $M$. Following Ref.\cite{Ziegler2017}, we will model the isolated impurities themselves as  $\delta$-function potentials,  
\begin{align}\label{EqHam96}
V({\bf r})=\sum_{l=1}^N U_l\ \delta({\bf r}-{\bf r}_l),
\end{align}
where $N$ is the number of impurities, and $U_l$ their strength. 
This ansatz models a situation in which the physical range, $\xi$, of individual potentials is comparable to $M^{-1}$ (such that they look spatially structureless from the perspective of the regularized Weyl Hamiltonian), yet larger than the momentum space separation $q_{\text{node}}^{-1}\gg M$ between different Weyl cones (such that we are still dealing with individual, uncoupled nodes.)

\subsection{Scattering free energy functional}

The advantage gained for our simplistic modeling of the impurities is that $T$-matrix methods can be applied to a straightforward computation of the Green's function and the DoS. Starting with the definition of the full and bare Green's function
\begin{align}\label{EqHam97}
G_0^{-1}=\omega^+-\hat{H}_0, \ G^{-1}=\omega^+-\hat{H}_0-\hat{V},
\end{align}
we represent the DoS as
\begin{align}\label{EqHam98}
\delta\nu(\omega)=-\frac{1}{\pi}\text{Im tr}\left(G-G_0\right)=-\frac{1}{\pi}\text{Im tr}\left(G_0\frac{\hat{V}G_0}{1-\hat{V}G_0}\right).
\end{align}
Now define the diagonal  $N\times N$ matrix $\hat{U}=\text{diag}(U_1, U_2, ... U_N)$,  and the projector on the impurity positions $\hat{P}=\text{diag}(|{\bf r}_1\rangle\langle{\bf r}_1|,|{\bf r}_2\rangle\langle{\bf r}_2|,...|{\bf r}_N\rangle\langle{\bf r}_N|)$. With these objects we simplify Eq.~\eqref{EqHam99} as 
\begin{align}\label{EqHam99}
\delta\nu(\omega)&=-\frac{1}{\pi}\text{Im tr}\left(\frac{\hat{U}}{1-\hat{U}\hat{P}G_0\hat{P}}G_0^2\hat{P}\right)=\frac{1}{\pi}\text{Im tr}\left(\frac{\partial_\omega \hat{G_0}
}{\hat{U}^{-1}-\hat{G_0}}\right)\nonumber\\
&=-\frac{1}{\pi}\partial_\omega \text{Im}\log\det\left(\hat{U}^{-1}-\hat{G_0}\right)\equiv-\frac{1}{\pi}\partial_\omega \text{Im} F,
\end{align}
i.e. the DoS is now represented as the derivative of a `free energy', $F=\log\det\left(\hat{U}^{-1}-\hat{G_0}\right)$, obtained by taking the determinant of an $2N\times 2N$ dimensional matrix. Here, $\hat{G}_0=\hat{P}G_0\hat{P}$, and we used that 
$G_0^2=-\partial_\omega G_0$. The $2\times 2$ block matrix elements entering the computation of $F$ are given by  $(\hat{G}_0)_{ij}=G_0({\bf r}_i-{\bf r}_j)$.

\begin{figure*}
    \includegraphics[width=0.7\linewidth]{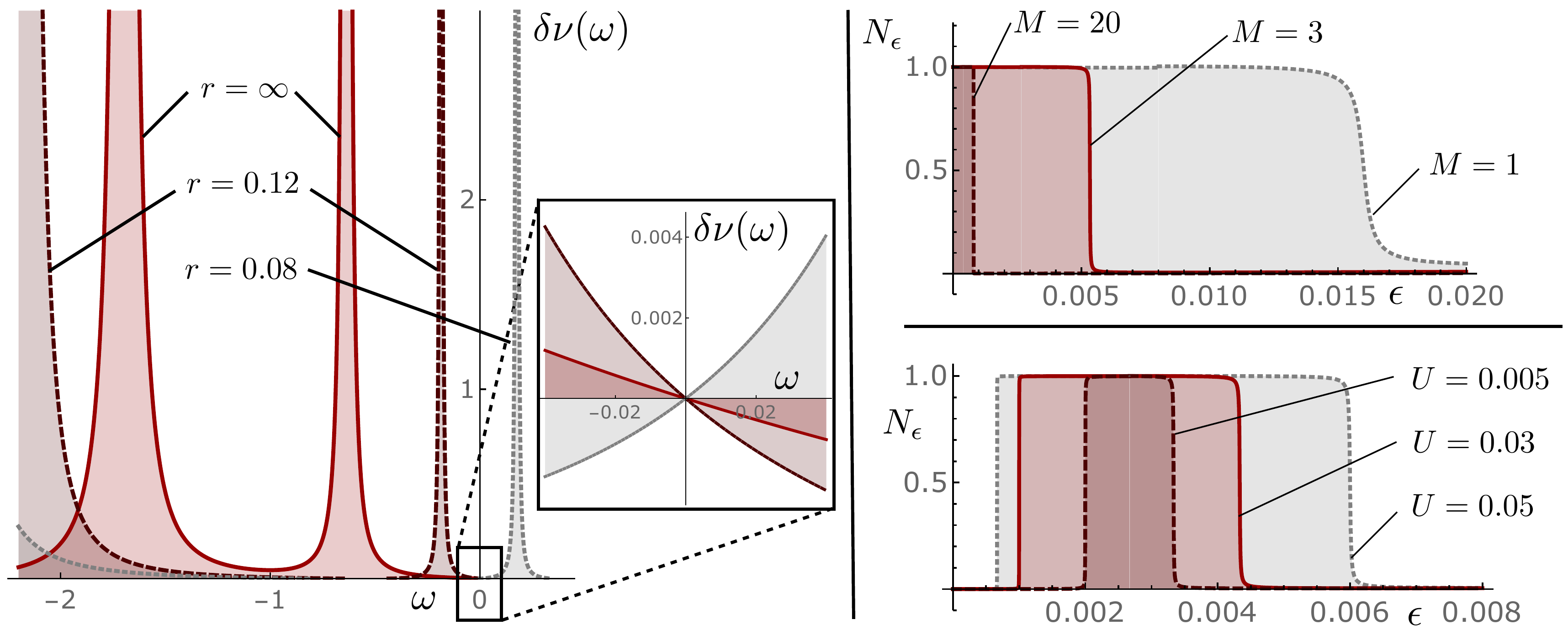}
    \caption{Left: Modification of the density of state
    $\delta\nu(\omega)$ by a pair of impurities $(U_1, U_2)=(0.02, 0.0075)/\pi$ at
    $M=100$. For infinite separation in space, $r=\infty$, the peaks correspond to
    the resonances of the bare impurities $\omega_{1,2}=1/(MU_{1,2})$. For smaller
    separation the resonances hybridize according to Eq.~\eqref{EqHam108}. The inset
    shows the vanishing DoS at $\omega=0$ and its $\sim\omega$ growth/decay. Right:
    Number of states $N_{\epsilon}$ accumulated in the frequency regime
    $\omega\in [-\epsilon, \epsilon]$ by the impurity potential. Top:
    $N_{\epsilon}$ for a resonance at $\omega=0$, i.e. $U_1=U_2=8\cdot
    10^{-3}$ and $r^2=U_1$ but different $M$. States are accumulated at $\omega=0$  but screened by a negative DoS in the vicinity of $|\omega|\sim 1/M$ such that $N_\epsilon\rightarrow0$ for $\epsilon>1/M$.  Bottom: For different resonances at $\omega>0$ set
    by $U_1=U_2=U\neq r^2$ and $M=3$. Around zero frequency no spectral weight is
    accumulated and a minimal energy $\omega>0$ is required to observe the effect of
    the impurity potential.}
    \label{fig4}
\end{figure*}

The matrix $\hat G_0$ contains the diagonal elements $(\hat G_0)_{ii}=G_0({\bf 0})$, which are singular in the $M\to \infty$ limit. (We here see how the $\delta$-impurity modeling brings about the need for regularization.)  
Switching to a Fourier representation, $G_0(\mathbf{r})= \int \frac{d^3k}{(2\pi)^3} G_0({\bf k})e^{i k_i r^i}$,
the momentum  space representation of the linear Weyl Hamiltonian,
\begin{align}\label{EqHam101}
    G_0(\mathbf{k})=(\omega^+-k_i \sigma^i)^{-1}=\frac{\omega^++k_i \sigma^i}{\omega^{+ 2}-k^2},
\end{align}
leads to an UV singularity at $\mathbf{r}=0$. We repair this by a 
 Pauli-Villars inspired regularization, in which the denominator is replaced by
\begin{align}
   \frac{1}{\omega^{+ 2}-k^2}\rightarrow\frac{M^2}{(\omega^{+ 2}-k^2)(M^2+k^2)},
\end{align}
with a large `mass', $M$. For small $k$, the additional factor can be ignored, and for large $k$ it removes the singularity by introducing a faster decay with an effective band curvature $M$.
The above integral may now be evaluated as (with $r=|{\bf r}|$)
\begin{align}\label{EqHam103}
    G_0({\bf r})=(\omega^+-i \sigma^i\partial_i)\frac{(-1)}{r}\frac{1}{4\pi}\left(e^{i\omega^+ r}-e^{-M r}\right).
\end{align}
The cases $r\not=0$ and $r=0$ require separate treatment. For $r>0$ but small energies, $\omega r<1$ we can forget about $\exp(-Mr)$, and  expansion in $\omega r$ yields
\begin{align}\label{EqHam104}
 G_0({\bf r})\overset{r>0}{=} \frac{(i\partial_{r^i} \sigma^i-\omega^+)}{4\pi}\frac{e^{i\omega^+r}}{r}=-\frac{\omega^+r+i n^i \sigma_i}{4\pi r^2 }+O(\omega^2),\ \ \ \ \ \ \ \ \end{align}
where $r^i = n^i r $.
For $r=0$, the introduction of $M$ does not suffice to regularize the superficially divergent $\sigma_i$-contribution to the integral. At the same time, this term vanishes by symmetry, and will disappear under stronger regularization schemes (as provided by any realistic lattice dispersion). The correct 
interpretation therefore is to ignore this term, which gives 
\begin{align}\label{EqHam105}
     G_0(\textbf{0})= -\frac{M+i\omega^+}{4\pi}\omega^++O(\omega^2). 
\end{align}

\subsection{Density of states in the presence of two impurities}
In the following we apply the above results to analyze the DoS of a system of just
two impurities,  $U_1, U_2$. This simple yet instructive case illustrates suffices to illustrate the essential features of more complex multi-impurity systems. We choose coordinates such that  
 ${\bf r}_1=(0,0,0)^T$ and ${\bf r}_2=(0,0,r)^T$. Using the above formulas for the Green's  functions, the free energy becomes the determinant of a simple $4\times 4$ matrix, and a straightforward computation yields
\begin{align}
    F&=-2\ln\left((U^{-1}_1+\tilde{M}\omega^+)(U^{-1}_2+\tilde{M}\omega^+)-r^{-4}-r^{-2}\omega^{+2}\right).
\end{align}
Here, we rescaled $U_{1,2}\to4\pi U_{1,2} $ for notational simplicity and defined
$\tilde{M}=M+i\omega^+$. 
Carrying out the differentiation, we obtain the result
\begin{align}
    \delta\nu(\omega)=-\frac{2}{\pi}\mathrm{Im}\frac{\tilde{M}(U^{-1}_1+U^{-1}_2+2\tilde{M}\omega^+)-2r^{-2}\omega^+}{(U^{-1}_1+\tilde{M}\omega^+)(U^{-1}_2+\tilde{M}\omega^+)-r^{-4}-r^{-2}\omega^{+2}}.
\end{align}
Inspection of the denominator in the limit $M\gg r, \omega$ shows that $\delta\nu$ has resonances at 
\begin{align}\label{EqHam108}
\omega_{\pm}=-\frac{U_1+U_2\pm\sqrt{(U_1-U_2)^2+4U_1^2U_2^2r^{-4}}}{2MU_1U_2}.
\end{align}
For a large spatial separation of the impurities $r\gg \sqrt{U_1U_2}$, the two impurities decouple and the resonances approach the values for single, individual impurities, $\omega_{\pm}=-(MU_{1,2})^{-1}$. However,  for diminishing  spatial separation we observe hybridization and a shifting of the isolated resonance frequencies, see Fig.~\ref{fig4}. 

For generic parameter values, an expansion of the DoS near zero yields (see Fig.~\ref{fig4}, inset)
\begin{align}
\delta\nu(\omega)=\frac{4r^4(U_1+U_2)}{\pi(r^4-U_1U_2)}\omega+O(\omega^2),
\end{align}
and $\delta\nu(0)=0$. However for the fine tuned configurations 
$U_1U_2=r^4$, the resonance centers shift to $\omega=0$ and the DoS near zero becomes singular (as evidenced by the diverging derivative in the linearization above.) 
for small $\omega$. Physically, this singularity reflects the formation of a zero energy bound state between two impurities. The pole-like nature of the DoS in this limit shows that this really is a state, as opposed to the resonances centered at finite energies for other parameter values. This state is the two-impurity analog of the states previously seen in the box potential. 

Notice that for resonances forming at a small energy $\omega_\mathrm{r}=\omega_{\pm}$ the DoS shows a peaks whose height diverges as $1/M \omega_\mathrm{r}^2$ while the width $\sim M \omega_\mathrm{r}^2$ shrinks in the inverse of the same parameter. Rather than focusing on the singularity of the DoS at $\omega_0\to 0$ a more rewarding approach is to investigate the number of states
\begin{align}
N_\epsilon=\int_{-\epsilon}^{+\epsilon}\delta\nu(\omega)d\omega=-\frac{1}{\pi} \left. \text{Im}F\right|_{-\epsilon}^{+\epsilon},
\end{align}
accumulated  inside a window of width $2\epsilon$ centered around zero energy. The upper panel on the right of Fig.~\ref{fig4} shows $N_\epsilon$ in the singular case $\omega_{\mathrm{r}}=0$, of a resonance tuned to zero energy. The figure shows that a positive spectral weight in a narrow region  $M \epsilon<U_1^{-1}+U_2^{-1}$ is balanced by negative spectral weight concentrated in a window roughly twice as large. In total no states are contained in windows exceeding this width. In the limit of a  perfectly linear Weyl dispersion $M^{-1}\rightarrow0$ no spectral weight is contained near zero energy. The bottom panel shows $N_\epsilon$ for resonances at generic energies, $\omega_r>0$. In this case, too, no weight is sitting near zero energy. The width $\epsilon$ needs to exceed $\omega_r$, in order to capture the resonance and induce a non-vanishing number of states.

We finally note that the scaling of peak widths with inverse $M$  might be behind the 
result of finite spectral density $\delta\nu(\omega\sim0)>0$ in numerical
works with non-zero curvature\cite{Pixley2016a,Pixley2016b,Wilson2016,Justin2018} and
the absence thereof in approaches with a perfectly linear
dispersion\cite{Sbierski2015,Sbierski2014,Kobayashi2014}. No matter how large $M$ our results imply the absence of spectral density \emph{at} zero energy. (As discussed in section~\ref{sub:statistical_distribution_of_the_dos}, zero width peaks at zero energy have zero statistical measure.) However, the increasing width of peaks for smaller $M$ makes it relatively easier to observe spectral weight in the immediate neighborhood of zero, and in a finite resolution numerical experiment this might be confused for DoS at zero. {\color{black} We also note that the resonance frequencies of our impurities decrease upon increasing the impurity strength at fixed band curvature and impurity position, cf. Eq.~\eqref{EqHam108}. In a finite resolution numerical investigation this would be indistinguishable from an increase of the zero energy DoS with increasing disorder strength. The latter phenomenon is observed in numerics, but from our current perspective might reflect a shift of fine resonances towards zero rather than DoS at zero.}

\section{Conclusion} 
\label{sec:Conclusion}
In this work, we have analyzed the spectral density of weakly disordered Weyl semimetals from three different perspectives. Our approaches were based on different modelings of the impurity potentials corresponding to statistically rare fluctuations, and on different analytic techniques. However, they all had in common that the DoS at zero energy remained vanishing and the nodal points remained preserved. 

To summarize the main characteristics, the simplistic box-function model afforded a full analytic solution of the problem, including a description of the statistical distributions generated by an ensemble of potentials of varying width and depth. We saw that the formation of resonances of increasing amplitude and sharpness (all the while at vanishing nodal density of states) implied singularities in the distribution. Within the statistical reading, the presence of an infinitely sharp resonance infinitely close to zero represents an event of measure zero and is not observed in any individual sample.

The second model replaced the boxes by a distribution of generic (Gaussian distributed) potentials. No rigorous analytic solution was possible, and instead we applied the standard, yet technically involved, method of large deviations and instanton calculus. Due to the necessary approximations only an upper limit for the DoS at zero energy double exponential in the weak disorder concentration could be obtained (while at any non-zero energy, the DoS came out finite). The approximations had to do with the presence of band curvature and inter-impurity correlations, which both could not be rigorously accounted for.

In this regard, the third model, a system of point like impurities, played a complementary role. Here, impurity correlations are of paramount importance, and band curvature was even necessary as a regularizing instance. The simple modeling of individual impurities as $\delta$-functions made an full solution possible. Again, the DoS at zero energy turned  out to be protected.

All three models predict a vanishing nodal DoS along with finite DoS in any
neighborhood of zero. The two phenomena coexist by virtue of ever more singular
resonant structures as zero is approached. We reasoned that the diminishing widths of such peaks reduces their statistical measure in the way that the limit --- a zero width peak of infinite height in a zero neighborhood of zero has zero measure and remains unobservable. However, in numerical experiments of finite resolution, this mathematical zero might be difficult to establish. This applies in particular to lattice simulations with finite band curvature (we saw that the width of resonant peaks in multi-impurity systems increases in the curvature parameter.) And of course the DoS
will become finite for potentials sharp enough to scatter between different Weyl
nodes. These real life intrusions imply that in numerical simulations an effectively finite DoS may be
observed, and the same goes for real experiments. In this regard, the protection
mechanisms discussed above below play more of a fundamental than an applied role.
Specifically, they show that the nodal DoS does qualify as an order parameter for a
quantum phase transition separating a weak from a strong disorder phase in the Weyl
system.

{\color{black} Finally, let us briefly address the role of  types of disorder, not
considered in this paper. We have already mentioned that disorder coupling different
Weyl nodes will completely change the picture. The coupling between sectors of
opposite chirality effectively removes the protection mechanism of the nodal
structures and renders the zero DoS finite. However, even for individual nodes more
general forms than the scalar potential of~\eqref{EqHam1} are conceivable: one
may include magnetic impurity scattering\cite{Roy2018} to induce random coupling proportional
to the Pauli matrices $\sigma_i$. In the semiclassical limit of a smooth impurity
potential this becomes equivalent to random shifts in the position of Weyl nodes in
the Brillouin zone. Such shifts, although still RG irrelevant, may
affect the system on different levels than the scalar disorder\cite{Volovik1996,Volovik2018}~ \footnote{G.~E. Volovik, private communication}. However, this physics is beyond the scope of
the present analysis, which hence remains limited to the case of non-magnetic
impurity scattering. }

\acknowledgments  We want to thank P.~W.~Brouwer, V.~Gurarie, R.~Nandkishore, L.~Radzihovski, G.~Refael, B.~Sbierski, G.~Volovik, J.~H.~Wilson, K.~Ziegler, and M.~Zirnbauer for fruitful discussions. This work has been supported by the German Research
Foundation (DFG) through CRC/TR 183 -- Entangled states of matter (project A02) and
the Institutional Strategy of the University of Cologne within the German Excellence
Initiative (ZUK 81). M.~B. thanks the Alexander von Humboldt
foundation for support.

\appendix
\section{Relation between scattering phase shift and density of states}
Consider a scattering problem with Hamiltonian $\hat{H}=\hat{H}_0+\hat{V}$, where $\hat{H}_0$ is the free Hamiltonian. The modification of the density of states by $\hat{V}$ can be expressed as
\begin{align}\label{EqHam79}
\delta\nu(\omega)&\equiv \nu(\omega)-\nu_0(\omega)\\&=-\frac{1}{\pi}\text{Im Tr}\left(\frac{1}{\omega^+-\hat{H}_0-\hat{V}}-\frac{1}{\omega^+-\hat{H}_0}\right),\nonumber
\end{align}
where $\omega^\pm=\omega\pm i0^+$ and $\nu(\omega)$ is the DoS of the scattering Hamiltonian $\hat{H}$ while $\nu_0(\omega)$ is the DoS of the free Hamiltonian. The trace is performed over the entire Hilbert space.

Exploiting the linearity of the trace and the definition for the imaginary part, one may rewrite Eq.~\eqref{EqHam79} such that~\cite{Langer61}
\begin{align}\label{EqHam80}
\delta\nu(\omega)&=\frac{1}{\pi}\partial_\omega {\rm{Im\ Tr}}\ln \left(1-\frac{1}{\omega^+-\hat{H}_0}\hat{V}\right)\\
&=\frac{1}{2\pi i}\partial_\omega{\rm{Tr}}\left[\ln\left(1-\frac{1}{\omega^+-\hat{H}_0}\hat{V}\right)-\ln\left(1-\frac{1}{\omega^--\hat{H}_0}\hat{V}\right)\right]\nonumber.
\end{align}
The second logarithm can be rewritten as
\begin{align}\label{EqHam81}
1-\frac{1}{\omega^+-\hat{H}_0}\hat{V}=1+2\pi i\delta(\omega-\hat{H}_0)\hat{V}-\frac{1}{\omega^--\hat{H}_0}\hat{V}.
\end{align}
Due to the $\delta$-function and the fact that $\omega^+$ has been eliminated from the expression, the imaginary part in $\omega^-$ can be dropped. This yields
\begin{align}\label{EqHam82}
\delta\nu(\omega)=\frac{1}{2\pi i}\partial_\omega{\rm{Tr}}\ln\Big(1+2\pi i\delta(\omega-\hat{H}_0)\hat{T}_{\omega}\Big),
\end{align}
where $\hat{T}_{\omega}=\hat{V}+\hat V\frac{1}{\omega-\hat{H}_0-\hat V}\hat{V}$ is the well-known $T$-matrix of scattering theory. Evaluating the trace in the basis of the eigenfunctions of the free Hamiltonian $\text{Tr}(...)=\int_{-\infty}^\infty d\omega\sum_{\kappa=1}^\infty\sum_{m_j=-\kappa+1/2}^{\kappa-1/2}\langle\omega, \kappa, m_j|...|\omega, \kappa, m_j\rangle$, the $\delta$-function constrains the energy integral to $\omega=\omega$ (otherwise the logarithm of unity vanishes). This reduces the trace to the on-shell scattering matrix $\hat{S}=1+2i\hat{T}$  in the eigenspace of states  with energy $\omega=\omega$. The unit-modular eigenvalues of the $S$-matrix define the scattering phase shifts through $e^{2i\delta}$, where $\delta$ is labeled by total angular momentum $\kappa$ and its $z$-component $m_j$.
For fixed $\kappa$, the phase shift is independent of $m_j$ and $2\kappa$-fold degenerate, which yields
\begin{align}
\delta\nu(\omega)&=\frac{1}{2\pi i}\partial_\omega \sum_{\kappa, m_j}\ln e^{2i\delta_{\kappa}(\omega)}=\frac{2}{\pi}\sum_{\kappa=1}^\infty\kappa\partial_{\omega}\delta_{\kappa}(\omega).\label{EqHam83}
\end{align}
This is the direct relation between the scattering phase shift and the DoS in the presence of a scattering potential $\hat{V}$. Which we will exploit below in order to discuss the DoS of a Weyl particle in the presence of a spherical box potential.

\section{Estimating the integral kernel of the massive modes}\label{AppIntkern}
In the main text, we realized the importance of the argument $\sin\theta e^{Q(r)}$ for the zero modes. In order to estimate the integral kernel \eqref{EqHam56}, it is convenient to perform a change of variables $(r,\theta)\rightarrow (r,y)$ with
\begin{align}\label{EqHam57}
   y\equiv \sin\theta\frac{r }{1+r^2}=\lim_{r^{\pm1}\rightarrow \infty}\sin\theta e^{Q(r)}.
\end{align}
It faithfully models the asymptotic behavior of $Q(r)$ at large and small distances, see Eq.~\eqref{EqHam51} (The azimuthal variable is inessential in the present context.) One may directly show that the measure transforms as
\begin{align}
 dV=dr d\theta\   r^2 \sin\theta =  \frac{(1+r^2)^2y}{\left(1-\left(\frac{y(1+r^2)}{r}\right)^2\right)^{1/2}} dr dy.
\end{align}
The integration limits are such that $y(r,\theta)$ ranges between $0=y(0,\theta)=y(\infty,\theta)=y(r,0)$ and a maximal value $1/2=y(1,\pi/2)$. For fixed $y$ the variable $r(y,\theta)$ assumes values between $r_\mp \equiv \frac{1}{2y}\mp \left( \frac{1}{(2y)^2}-1 \right)^{1/2} $, where for $y\ll1$,  $r_-\simeq 2y\ll 1$ and $r_+\simeq y^{-1}$. Functions with support far outside the instanton correlation range, $r\gg 1$, have support in a narrow range of $y$ close to zero.

We aim to define a class of zero mode functions $\{F_{a}\}$ orthogonal relative to the measure $dV |\varphi_I|^2=dV \frac{1}{(1+r^2)^2}$, where $\varphi_I$ are the instanton wave functions. According to Eq.~\eqref{EqHam57}, choosing a set of functions that has support only at $r\rightarrow\infty,0$, $F_{a}(r,\theta)\equiv F_a(y)$, which leads to the condition
\begin{align}
    \langle F_a,F_b\rangle \equiv \frac{\pi}{2} \int_0^{1/2}dy  \bar F_a(y) F_b(y)=\delta_{ab}.\label{Norm}
 \end{align}
Here the $r$-integration is independent of $y$ and
\begin{align}\label{EqHam60}
 \int_{r_-(y)}^{r_+(y)}
     ydr\left(1-\left(\frac{y (1+r^2)}{r}\right)^2\right)^{-\frac{1}{2}}=\frac{\pi}{2}.
\end{align}
 
In the regime of interest, $y\ll 1$, one finds $\mathcal{V}^{-1}(r,y)\simeq \frac{r}{2}\left( 1+\frac{3}{2(1+r^2)}+\frac{7}{2}y^2(1+r^2) \right)$.
This allows us to rewrite the integral kernel in Eq.~\eqref{EqHam56} and determine an upper bound for $\delta S$ as
\begin{align}\label{EqHam61}
     I_{a,b}&=\int dV |\varphi|^2 \mathcal{V}^{-1}\partial_z \bar F_a \partial_z F_b\nonumber\\
     &=\int_0^{1/2}ydy \int_{r_-(y)}^{r_+(y)}
     \frac{\mathcal{V}^{-1} dr}{\left(1-\left(\frac{y (1+r^2)}{r}\right)^2\right)^{1/2}}\partial_z \bar F_a(y) \partial_z F_b(y)\nonumber\\
     &=2\int_0^{1/2}dy\ y^3 \bar F_a'(y)F_b'(y)\underbrace{\int_{r_-(y)}^{r_+(y)}\frac{2r \left(\frac{r^2}{(1+r^2)^2}-y^2\right)^{\frac{1}{2}}}{\left(1+r^2\right)}\mathcal{V}^{-1} dr}_{\le -\ln(y/2)}\nonumber\\
&=\le-2\int_0^{1/2}dy y^3 \ln(y/2)\,\bar F_a'(y)F_b'(y).
\end{align}
In the third line above, we have evaluated the derivatives as
\begin{align}\label{EqHam62}
\partial_zF_a\partial_zF_b=F_a'F_b' \left(\frac{\partial y}{\partial z}\right)^2=F_a'F_b'\,4y^2\left(\frac{r^2}{(1+r^2)^2}-y^2\right).\ \ \ 
\end{align}
Observe that the Jacobian factor vanishes both for small and large $r$. For small $r$, this reflects the independence of  $F_a(y)\simeq F_a(r\sin \theta)$, on the $z$-differentiation coordinate. For large $r$ the entire space-volume is compressed into a narrow region of $y$-coordinates, and the Jacobian accounts for this volume distortion.

\section{Integration measure of the zero mode integral}
\label{IntMeas}
In contrast to linear field transformations $\varphi\rightarrow\varphi_I+\varphi$, which leave the integration measure unmodified, the unitary rotations $\varphi\rightarrow U\varphi$ are nonlinear in the Grassmann fields and introduce a nontrivial transformation of the measure $\mathcal{D}[\psi,\bar\psi]\rightarrow \mathcal{D}[U, \bar U, \varphi, \bar\varphi]$. In the main text we have focussed on modification of the action $S_{\text{eff}}$ by $U$. In this section, we show that the integration measure remains well-defined under all transformations, i.e. neither does it feature any singularities nor does it depend on the Grassmann zero modes. This demonstrates the validity of our analysis and, in combination with the main text, completes the instanton analysis of the Weyl semimetal with Gaussian disorder.

As with conventional Riemannian manifolds, the integration measure for integration over super-manifolds can be obtained by inspection of the appropriate `metric', $d\Xi^T G d\Xi$, where $\Xi$ is a supervector of integration variables, and $G$ a super-matrix playing the role of a metric tensor. The Jacobian is then obtained as $J=\sqrt{\mathrm{sdet}(G)}$, where $\mathrm{sdet}(G)=\det(G_{cc})/\det(G_{aa}-G_{ac}G_{cc}^{-1}G_{ca})$ is the super-determinant expressed via the commuting $G_{cc,aa}$ and anti-commuting $G_{ac,ca}$ blocks of $G$ \cite{EfetovBook,Berezin} (matrix elements between a commuting ($c$) and an anti-commuting $(a)$ field must be anti-commuting).  

Presently, the starting point is $\int dx d\bar \psi d\psi=\int dx(d\bar \phi d\phi + d\bar \chi d\chi)$, where $G=\mathds{1}$ corresponds to a flat integration measure, $J=1$. We now substitute $\psi=U(\varphi,0)^T$, where $\varphi$ is commuting, and $U=W\cdot V$ the Grassmann rotation defined in Eq.~\eqref{EqHam35}. The chain rule yields $d\psi=dU(\varphi,0)+U(d\varphi,0)$. The bilinear form then reads as $
d\bar\psi d\psi= d\bar \varphi d\varphi+ d\bar \varphi (U^{-1}dU)_{cc}\varphi+ \bar \varphi (dU^{-1} U)_{cc} \varphi+ \bar \varphi (dU^{-1} dU)_{cc}\varphi)$ where $(... )_{cc}$ indicates the projection onto the bosonic sector. Evaluation of the differentials yields the matrices
\begin{align}
G_{cc}&=\partial_{d\varphi}d\bar\psi d\psi\overset{\leftarrow}{\partial}_{d\varphi}=\mathds{1},\\
G_{ac}&=\partial_{d\Delta,d\nu}d\bar\psi d\psi\overset{\leftarrow}{\partial}_{d\varphi}=-G_{ca}^T\\
&=\frac{1}{2}\left(\begin{array}{cccc}-\Delta\bar\phi_1&-\bar\Delta\bar\phi_1&-2\Delta\bar\phi_1-\nu\bar\phi_1&-2\bar\Delta\bar\phi_1-\bar\nu\bar\phi_1\\ -\Delta\bar\phi_2&-\bar\Delta\bar\phi_2&2\Delta\bar\phi_2-\nu\bar\phi_2&2\bar\Delta\bar\phi_2-\bar\nu\bar\phi_2\nonumber\\
\Delta\phi_1&\bar\Delta\phi_1&2\Delta\phi_1+\nu\phi_1&2\bar\Delta\phi_1+\bar\nu\phi_1\\
\Delta\phi_2&\bar\Delta\phi_2&-2\Delta\phi_2+\nu\phi_2&-2\bar\Delta\phi_2+\bar\nu\phi_2 \end{array}\right),\nonumber\\
G_{aa}&=\partial_{d\Delta,d\nu}d\bar\psi d\psi\overset{\leftarrow}{\partial}_{d\Delta,d\nu}.
\end{align}
Here the gradient terms $\partial_{d\varphi}\equiv(\partial_{d\varphi_1},\partial_{d\varphi_2},\partial_{d\bar\varphi_1},\partial_{d\bar\varphi_1})^T$ and $\partial_{d\Delta,d\nu}\equiv(\partial_{d\bar\Delta},\partial_{d\Delta},\partial_{d\bar\nu},\partial_{d\nu})^T$ act from the left and their transposes $\overset{\leftarrow}{\partial}_{d\varphi}, \overset{\leftarrow}{\partial}_{d\Delta, d\nu}$ act from the right. The expression for $G_{aa}$ is rather lengthy but obtained along the lines of $G_{cc}, G_{ca}, G_{ac}$. 
Due to the product form of $U$, the fermion-fermion and boson-fermion sector individually still contain Grassmann variables but the difference 
\begin{align}
&G_{aa}-G_{ac}G_{cc}^{-1}G_{ca}=\frac{i}{2}\sigma_y\otimes\left(\begin{array}{cc}|\varphi|^2 &\bar\varphi\sigma_z\varphi\\ \bar\varphi\sigma_z\varphi & |\varphi|^2\end{array}\right)&
\end{align}
is a matrix with real coefficients. Since the transformation does not affect the purely commuting sector, its determinant remains $\det (G_{cc})=1$. This leads to  $J=4\prod_x(|\varphi_x|^4-(\bar\varphi_x\sigma_z\varphi_x)^2)^{-1}=\prod_{i,x}| \varphi_{i,x}|^{-2}$.

In a final step, we turn to the Weyl spin center coordinates in the Grassmann sector, $d\bar \eta_x d\eta_x$ and expand the isotropic fields $\eta_x$ in a basis of functions $\{F_a\}$ orthogonal
relative to the scalar product $\langle  F_a,F_b\rangle\equiv \int_x \, \bar F_a \, g
\,F_b=\delta_{ab}$, with diagonal metric $g_x=
|\varphi_{x}|^2$ introduced in Eq.~\eqref{Norm}. The expansion $\eta_x=\sum_a F_a(x) \xi_{a}$,  introduces the
functional determinant, $\prod_{x}d \bar \eta_{x} \eta_{x}=\prod_{a}d \bar
\xi_{a} \xi_{a} \left|\det(F_{a}(x))\right|^{-2}$. 
This logic can be again applied to the massive fluctuations $\Delta$, which yields an additional determinant factor $\left|\det(F_{a}(x))\right|^{-2}$ and, as a consequence  of the normalization, regularizes the Jacobian.

\bibliography{Diss}
\end{document}